\documentclass[10pt]{article}
\pdfoutput=1
\usepackage{epsfig}
\usepackage{amsfonts}
\usepackage{mathrsfs}
\usepackage{latexsym}
\usepackage{color}
\usepackage{amsmath,amssymb}
\usepackage{verbatim}
\definecolor{red}{rgb}{1,0,0}
\usepackage{cite}
\numberwithin{equation}{section}

\newcommand{\Riem}[4]{R_{#1 #2\phantom{#3}#4}^{\phantom{#1 #2}#3}}

\def\bea{\begin{eqnarray}} 
\def\eea{\end{eqnarray}}
\def\be{\begin{equation}} 
\def\ee{\end{equation}} 
\def\ba{\begin{array}}
\def\ea{\end{array}} 
\def\nn{\nonumber}

\begin{document}

\thispagestyle{empty}

\renewcommand{\thefootnote}{\fnsymbol{footnote}}
\setcounter{footnote}{0}
\setcounter{figure}{0}

\begin{center}

{\Large{\textbf{Splitting Ward identity}}}

\vspace{1cm}

{\large Mahmoud Safari}

\vspace{5mm}

\textit{School of Particles and Accelerators,\\ Institute for Research in Fundamental Sciences (IPM),
P.O.Box 19395-5531, Tehran, Iran,\\ E-mail: msafari@ipm.ir}

\vspace{5mm}

\begin{abstract}
Within the background-field framework we present a path integral derivation of the splitting Ward identity for the one-particle irreducible effective action in the presence of an infrared regulator, and make connection with earlier works on the subject. The approach is general in the sense that it does not rely on how the splitting is performed. This identity is then used to address the problem of background dependence of the effective action at an arbitrary energy scale. We next introduce the modified master equation and emphasize its role in constraining the effective action. Finally, application to general gauge theories within the geometric approach is discussed. 
\end{abstract}

\end{center}

\vspace{1cm}

\setcounter{page}{1}
\renewcommand{\thefootnote}{\arabic{footnote}}
\setcounter{footnote}{0}

\section{Introduction}
The notion of exact renormalization group originated from the pioneering work of Wilson \cite{wk}. Since then, it has re-emerged in various formulations \cite{wh,polchinski_rg,wetterich_rg,morris_rg}. Among them is the approach taken in \cite{wetterich_rg,morris_rg} where, contrary to \cite{wh,polchinski_rg} which study the scale dependence of the Wilsonian effective action, one deals with the scale dependence of the generator of one-particle irreducible diagrams, hereafter referred to simply as the effective action. The dependence on the energy scale $k$ is introduced by adding to the ultraviolet action an effective mass term for the dynamical fields, with a scale dependent mass $R_k(q^2)$, usually referred to as the cutoff kernel, which decreases monotonically with momentum $q$. This resembles Wilson's notion of incomplete integration, where the path integral over heavier modes is less suppressed.
It turns out that the scale dependence of this effective action $\Gamma_{\!k}$ is ruled by an equation which is exact, in the sense that it does not rely on the existence of any small expansion parameter.
This equation relates the scale derivative of the effective action to its second derivative with respect to the classical dynamical fields $\Gamma^{(2)}_{\!k}$
\be \label{frge}
\partial_t\Gamma_{\!k} = {\textstyle{\frac{1}{2}}}\mathrm{Tr}\big[(\Gamma^{(2)}_{\!k}\!\!+\! R_k)^{\!-\!1}\partial_t R_k\big]
\ee
where $t=\log k$. Despite being exact, finding solutions to this equation without resorting to any approximation seems out of reach. In practice one truncates the effective action to reduce the parameter space to a lower dimensional subspace, where the equations can be solved. This reduced parameter space can be finite or still infinite dimensional. For reviews on exact renormalization, and especially the approach of \cite{wetterich_rg,morris_rg}, see \cite{morris_review,bb_review,btw,polonyi_review,pawlowski_review,gies_review,rosten_review}.

Finding a consistent truncation requires additional care when using the background-field method  \cite{abbott_npb1981,abbott_app1982}. The background-field method is used widely in Yang-Mills theory and field theory of gravity for the computational facility and conceptual insight it provides. Its use is also necessary for the construction of a covariant effective action \cite{vilkovisky_gospel1984,vilkovisky_npb1984,dewitt_effact}, for both gauge and non-gauge theories. When using the background-field method, apart from the usual Lorentz symmetry and possible internal symmetries of the theory, there will be extra relations among the couplings of different operators in the effective action which originate from the fact that the background and quantum fields enter the ultraviolet action in a specific combination, namely the total field. These constraint relations are governed by some sort of Ward identities, which we generally refer to as splitting Ward identities. 

Considering nontrivial instances studied in the literature, such an identity was first used for the standard (infrared) effective action with nonlinear quantum-background split \cite{hps}, to prove renormalizability of general nonlinear sigma models in two dimensions. Within the renormalization group approach of \cite{wetterich_rg,morris_rg}, it has appeared in \cite{reuter_wetterich_npb1994,reuter_wetterich_9708,litim_pawlowski_9901,
litim_pawlowski_0203,litim_pawlowski_0208} and emphasized more recently in \cite{morris_1312,morris_1502}, for linear splitting of the field. For geometrical effective actions, which require a nonlinear split, it was first pointed out in \cite{pawlowski_0310,pawlowski_review}, in the context of gauge theories. Attempts to apply the equation in the case of nonlinear field splitting, were made in \cite{pawlowski_1203} for quantum gravity, in the geometric approach. 
 
In this work we give a general path integral derivation of the splitting Ward identities leading to the above mentioned constraint relations, in the presence of an infrared regulator, and show how these constraints are consistent with the functional flow equation of the effective action. Particular attention is paid to the choice of measure. Employing these identities, the problem of background dependence is then addressed for the most general case of quantum-background split, which, even in its special case of linear split, generalizes some recent results in the literature \cite{morris_1312,morris_1502}. We then introduce the modified master equation and explain how, with its aid, these identities can be applied in practice to put constraints on the form of the effective action. As a check, the modified master equation is explicitly shown to be satisfied at the one-loop level, irrespective of the scheme of regularization. All this is done without reference to a specific way of splitting the total field. Next, after reviewing the advantages of the exponential splitting and the notion of covariant effective action, we discuss general covariance of the results of earlier sections. Finally we explain how these results can be applied to gauge theories within the geometric approach. In this paper we set the framework and leave the application to a future publication. 

\section{Modified splitting Ward identity}  

\subsection{Setup and derivation of the identity} \label{derivation}

The starting point of our discussion is the quantization of a bare theory with action $S[\phi]$, for which the background-field method is to be employed. Here, in general $\phi$ is meant to denote a set of fields $\phi^i$, with $i$ regarded as a generalized index, including the label of fields, possible Lorentz indices, and also the spacetime/momentum argument. The fields are then chosen to be split into a background $\varphi^i$ and a fluctuation field $\xi^i$, so that $\phi^i(\varphi,\xi)$ is now a function of $\varphi^i$ and $\xi^i$, and such that $\phi^i(\varphi,0)=\varphi^i$. With foresight, the notation is chosen to match that commonly used for the exponential splitting, discussed in more detail in section \ref{cef} and appendix \ref{exp}, which is going to be our main application, but at this point we do not specify how the splitting is done. For simplicity of notation, throughout the paper, we use a dot to denote contraction of the generalized indices. The following formalism applies to non-gauge theories and gauge theories prior to gauge fixing. We defer a discussion of the gauge fixing procedure to section \ref{gauge}.

The generator of connected $n$-point functions $W_k[\varphi,J]$ is a functional of the background and a source field $J_i$ given by the path integral
\be  \label{W}
\exp\left(-W_k[\varphi,J]\right) = \int \!\! D\phi \;\mu(\phi) \,\; \mathrm{exp}\left(-S[\phi]-S_k[\varphi,\xi]- J\!\cdot\!\xi\right).
\ee
A dependence on the energy scale $k$ is introduced by adding a cutoff term, bilinear in the dynamical fields 
\be
S_k[\varphi,\xi] = {\textstyle{\frac{1}{2}}}\, \xi\!\cdot\! R_k(\varphi) \!\cdot\!\xi,
\ee
with a cutoff kernel which depends only on the background field, and vanishes at $k=0$. Also, the sole assumption on the integration measure is that it depends exclusively on the total field. This assumption is in fact not of central importance and is irrelevant in certain regularization schemes, as we will comment on later.
The cutoff and source terms break the single-field dependence of the exponent in \eqref{W}, and therefore lead to an (off-shell) effective action which, in principle, depends on how the total field is split. The scale dependent effective action is defined through the modified Legendre transform
\be \label{eaa_def}
\Gamma_k[\varphi,\bar\xi] = W_k[\varphi,J] - J\!\cdot\!\bar\xi -S_k[\varphi,\bar\xi],
\ee
where $\bar\xi^i=\langle\xi^i\rangle$ is the expectation value of $\xi^i$. 

The quantum-background split, being a field redefinition, does not affect physical quantities, but will have nontrivial consequences for the off-shell effective action. As emphasized in \eqref{eaa_def}, the effective action is no longer a function of a single field but depends separately on both the background and fluctuations. However, the fact that the bare action is a function of a single field must leave some trace on the form of the effective action. In terms of symmetries, the bare action is invariant under a set of simultaneous transformations of the background and the fluctuation field that leaves the total field unaltered 
\be  \label{symm}
\varphi^i \rightarrow \varphi^i + \delta\varphi^i, \hspace{5mm}
\xi^i \rightarrow \xi^i + \delta\xi^i; \hspace{1cm}
\phi^i(\varphi, \xi) \rightarrow \phi^i(\varphi + \delta\varphi, \xi + \delta\xi) = \phi^i(\varphi, \xi).
\ee
This symmetry will be inherited, possibly in a deformed way, by the effective action, which is manifested through the corresponding Ward identity, called the modified splitting Ward identity (mspWI), which we now wish to prove. The presence of an infrared regulator provides a modification to the analogue splitting Ward identity in the absence of this term, and hence the name `modified'. 

The derivation of the identity is rather straightforward. One varies eq.\eqref{W} with respect to $\varphi^i$. In doing so, one also varies the dummy variable $\xi^i$ such that the total field is left unchanged. This results in the following equality 
\be  \label{step1}
\frac{\delta W_k[\varphi,J]}{\delta \varphi}\!\cdot\!\delta\varphi = \left\langle \frac{\delta S_k[\varphi,\xi]}{\delta \varphi}\!\cdot\!\delta\varphi + \frac{\delta S_k[\varphi,\xi]}{\delta \xi}\!\cdot\!\delta\xi + J\!\cdot\!\delta\xi\right\rangle.
\ee
We remind here the formulas for the functional derivatives of the connected and one-particle irreducible correlation function generators which follow easily from  \eqref{W} and \eqref{eaa_def} 
\be  \label{step2}
\frac{\delta W_k[\varphi,J]}{\delta J_i} = \bar\xi^i, \hspace{1cm} 
\frac{\delta\Gamma_k[\varphi,\bar\xi]}{\delta\bar\xi^i} = -J_i - \frac{\delta S_k[\varphi,\bar\xi]}{\delta\bar\xi^i}, \hspace{1cm}
\frac{\delta \Gamma_k[\varphi,\bar\xi]}{\delta\varphi^i} = \frac{\delta W[\varphi,J]}{\delta \varphi^i}-\frac{\delta S_k[\varphi,\bar\xi]}{\delta \varphi^i}.
\ee
Using these identities, and shuffling terms a bit, eq.\eqref{step1} is rewritten as
\be  \label{step3}
\frac{\delta \Gamma_k[\varphi,\bar\xi]}{\delta\varphi}\!\cdot\!\delta\varphi +\frac{\delta\Gamma_k[\varphi,\bar\xi]}{\delta\bar\xi}\!\cdot\!\langle\delta\xi\rangle = \left\langle\frac{\delta S_k[\varphi,\xi]}{\delta \xi}\!\cdot\!\delta\xi\right\rangle -\frac{\delta S_k[\varphi,\bar\xi]}{\delta\bar\xi}\!\cdot\!\langle\delta\xi\rangle +\left\langle\frac{\delta S_k[\varphi,\xi]}{\delta \varphi}\right\rangle\!\cdot\!\delta\varphi -\frac{\delta S_k[\varphi,\bar\xi]}{\delta \varphi}\!\cdot\!\delta\varphi.
\ee
Dropping the arbitrary variation $\delta\varphi^i$, the last two terms on the right-hand side can be reorganized into an expression in terms of the connected two-point function, which is related to the one-particle irreducible two-point function using the first two equations in \eqref{step2} 
{\setlength\arraycolsep{2pt}
\bea
\left\langle\frac{\delta S_k[\varphi,\xi]}{\delta \varphi^i}\right\rangle -\frac{\delta S_k[\varphi,\bar\xi]}{\delta \varphi^i} &=& \frac{1}{2}\,\left\langle \xi\!\cdot\!\frac{\delta R_k(\varphi)}{\delta\varphi^i}\!\cdot\!\xi \right\rangle -\frac{1}{2}\, \bar\xi\!\cdot\!\frac{\delta R_k(\varphi)}{\delta\varphi^i}\!\cdot\!\bar\xi \nn\\
&=& -\frac{1}{2}\,\mathrm{Tr}\left[\frac{\delta^2 W_k[\varphi,J]}{\delta J\delta J}\,\frac{\delta R_k(\varphi)}{\delta\varphi^i}\right] = \frac{1}{2}\,\mathrm{Tr}\left[\left(\frac{\delta^2\Gamma_k[\varphi,\bar\xi]}{\delta\bar\xi\delta\bar\xi}+R_k(\varphi)\right)^{\!\!\!-\!1}\!\!\frac{\delta R_k(\varphi)}{\delta\varphi^i}\right].
\eea}%
The traces in the second line denote cyclic contraction of indices. Combining this result with eq.\eqref{step3}, we finally arrive at the mspWI 
\be  \label{mspwi} 
\boxed{\frac{\delta \Gamma_{\!k}[\varphi,\bar\xi]}{\delta\varphi^i} +\frac{\delta\Gamma_{\!k}[\varphi,\bar\xi]}{\delta\bar\xi}\!\cdot\!\left\langle\!\frac{\delta\xi}{\delta\varphi^i}\!\right\rangle \!-\! \frac{1}{2}\,\mathrm{Tr}\!\left[\!\left(\!\frac{\delta^2\Gamma_{\!k}[\varphi,\bar\xi]}{\delta\bar\xi\delta\bar\xi}\!+\!R_k(\varphi)\!\right)^{\!\!\!-\!1}\!\!\!\frac{\delta R_k(\varphi)}{\delta\varphi^i}\right]\!\! +\!\bar\xi\!\cdot\! R_k(\varphi)\!\cdot\!\left\langle\!\frac{\delta\xi}{\delta\varphi^i}\!\right\rangle \!\! - \!\left\langle \xi\!\cdot\! R_k(\varphi)\!\cdot\!\frac{\delta\xi}{\delta\varphi^i}\right\rangle =0}
\ee
Considering $\xi^j(\varphi,\phi)$ as a function of the background and the total field, the functional derivative in $\delta\xi^j/\delta\varphi^i$ is understood to be taken while keeping the total field fixed. 

Let us consider at this point some special cases of this identity. In the absence of a regulator $R_k(\varphi)=0$, the mspWI simplifies to an identity, similar in structure to the familiar splitting Ward identity for exponential quantum-background split \cite{hps,rebhan_npb1988,burgess_kunstatter,kunstatter_1992}
\be \label{spwi}
\frac{\delta \Gamma_{\!k}[\varphi,\bar\xi]}{\delta\varphi^i} +\frac{\delta\Gamma_{\!k}[\varphi,\bar\xi]}{\delta\bar\xi}\!\cdot\!\left\langle\!\frac{\delta\xi}{\delta\varphi^i}\!\right\rangle =0.
\ee
If in addition, $\delta\xi^i$ depends only on the background field or is at most linear in $\xi^i$, then we will have $\langle\delta\xi^i\rangle =\delta \langle\xi^i\rangle$ and the above equation translates to the fact that the symmetry of the bare action is also a symmetry of the effective action. Here, in the general case where $R_k(\varphi)$ is nonvanishing and $\delta\xi^i$ can have higher order terms in $\xi^i$, there will be modifications to this statement. 

Also, in the case of linear splitting $\phi^i = \varphi^i+ \xi^i$ we have $\delta\xi^i = -\delta\varphi^i$, so the last two terms in \eqref{mspwi} cancel out and the mspWI reduces to the modified \textit{shift} Ward identity  
\be  \label{mshwi} 
\frac{\delta \Gamma_{\!k}[\varphi,\bar\xi]}{\delta\varphi^i} -\frac{\delta\Gamma_{\!k}[\varphi,\bar\xi]}{\delta\bar\xi^i}- \frac{1}{2}\,\mathrm{Tr}\!\left[\left(\!\frac{\delta^2\Gamma_{\!k}[\varphi,\bar\xi]}{\delta\bar\xi\delta\bar\xi}\!+\!R_k(\varphi)\!\right)^{\!\!\!-\!1}\!\!\frac{\delta R_k(\varphi)}{\delta\varphi^i}\right] =0,
\ee
to which \cite{reuter_wetterich_npb1994,reuter_wetterich_9708,litim_pawlowski_9901,litim_pawlowski_0203,litim_pawlowski_0208} provide some early references. The significance of using this identity along with the flow equation has been stressed more recently in \cite{morris_1312,morris_1502}.

\subsection{Diagrams and shorthand notation} \label{short}

So far, we have used the notation in its expanded form to make the steps of the derivation clear. Now that we have finished this task, we introduce some shorthand notation which facilitate handling the equations significantly. From now on, for conciseness, we drop the index $k$ on $R_k$ and $\Gamma_k$, denote $\partial_t$ by an overdot when there is no ambiguity, and define for the functional $Q[\varphi,\xi]$
\be \label{der}
Q_{;\, i_1\cdots i_n},_{j_i\cdots j_m} \equiv \frac{\delta^{n+m} Q}{\delta\xi^{i_1}\cdots\delta\xi^{i_n} \delta\varphi^{j_1}\cdots \delta\varphi^{j_m}}.
\ee
In a more general sense, the notations `$,$' and `$;$' will also be used later to denote, respectively, partial derivatives with respect to the first and second arguments. Let us also denote the propagator and its inverse by $G^{ij}$ and $G_{ij}$ respectively 
\be 
G_{ij} = \Gamma\!_{;ij} + R_{ij}, \hspace{1cm} G^{ik}G_{kj} = \delta^i_j.
\ee
Using these compact notations, the mspWI \eqref{mspwi} is rewritten as
\be \label{mspwi_comp} 
\boxed{\mathcal{N}_i \equiv \Gamma\!,_{i} +\Gamma\!_{;j}\langle\xi^j\!\!,_{i} \rangle -{\textstyle{\frac{1}{2}}}\, G^{mn}(R_{nm}),_{i} +\bar\xi^m R_{mn}\langle\xi^n\!\!,_i\rangle - \left\langle \xi^m R_{mn} \xi^n\!\!,_i\right\rangle =0}
\ee
where the quantity $\mathcal{N}_i$, whose $k$-dependence is implicit, defines the expression on the left-hand side of this equation and is introduced for later use. 

In order to gain more insight into the mspWI, one can write it in a more explicit way, by expanding the quantity $\xi^j\!\!,_i$ in powers of the fluctuations, with background dependent coefficients
\be   \label{dxi} 
\xi^j\!\!,_i = \sum_{n=0}^\infty C^j_{i,i_1i_2\cdots i_n}\,\xi^{i_1}\xi^{i_2}\cdots\xi^{i_n}.
\ee 
These coefficients are taken to be symmetric in their lower indices $i_1i_2\cdots i_n$, without loss of generality. The expression for $\langle \xi^j\!\!,_i\rangle$ as a function of $\bar\xi^i$ is now seen more clearly in terms of diagrams. With the help of the Feynman rules defined in appendix \ref{feyn}, the second term on the left-hand side in eq.\eqref{mspwi_comp} is written in the following way 
{\setlength\arraycolsep{2pt}
\bea  
\Gamma\!_{;j}\langle\xi^j\!\!,_i\rangle  
&=& \sum_{n=0}^\infty \Gamma\!_{;j}\, C^j_{i,i_1i_2\cdots i_n}\, \langle\xi^{i_1}\xi^{i_2}\cdots\xi^{i_n}\rangle 
=  \; \sum\!\!\! \raisebox{-9mm}{\includegraphics[trim=7cm 22cm 7cm 2cm, clip=true, width=0.15\textwidth]{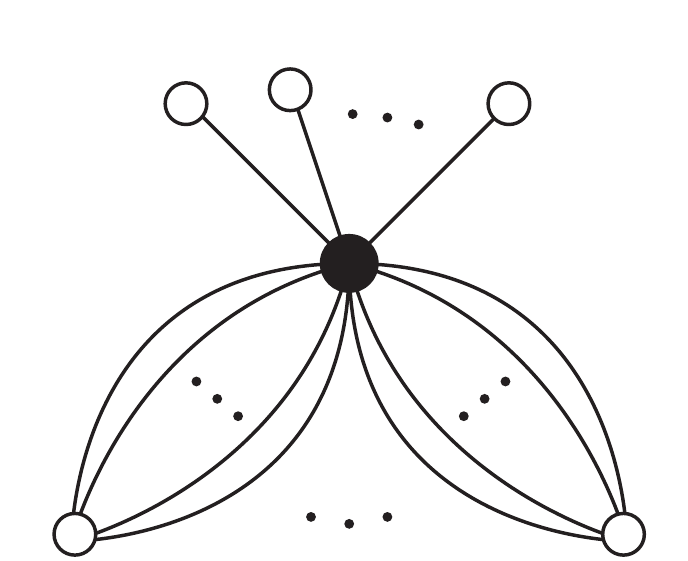}}  \label{d1}
\eea%
where the sum is over all possible diagrams of the given form, whose number is infinite, and include diagrams of all loop orders. The vertex with a black circle, whose index $i$ is implicit, represents $\Gamma\!_{;j}\, C^j_{i,i_1i_2\cdots i_n}$. The white circles are connected $n$-point functions of the fluctuation fields. In particular, the white circles connected by a single line are nothing but $\bar\xi$. The last two terms in \eqref{mspwi_comp} are also written as
{\setlength\arraycolsep{2pt}
\bea  
\bar\xi\!\cdot\! R\!\cdot\!\langle\xi,_i\rangle - \left\langle \xi\!\cdot\! R\!\cdot\!\xi,_i\right\rangle  
&=& R_{pq} \left(\langle \xi^q\!\!,_i \rangle \bar\xi^p - \langle \xi^q\!\!,_i \xi^p \rangle\right) \nn\\[2mm]
&=& \sum_{n=1}^\infty \, R_{pq}\, C^q_{i,i_1i_2\cdots i_n}\left(\langle\xi^{i_1}\xi^{i_2}\cdots\xi^{i_n}\rangle \bar\xi^p - \langle\xi^{i_1}\xi^{i_2}\cdots\xi^{i_n}\xi^p\rangle\right) 
\nn\\[-4mm]
&=&  \sum\!\! \raisebox{-9mm}{\includegraphics[trim=7cm 22cm 7cm 2cm, clip=true, width=0.15\textwidth]{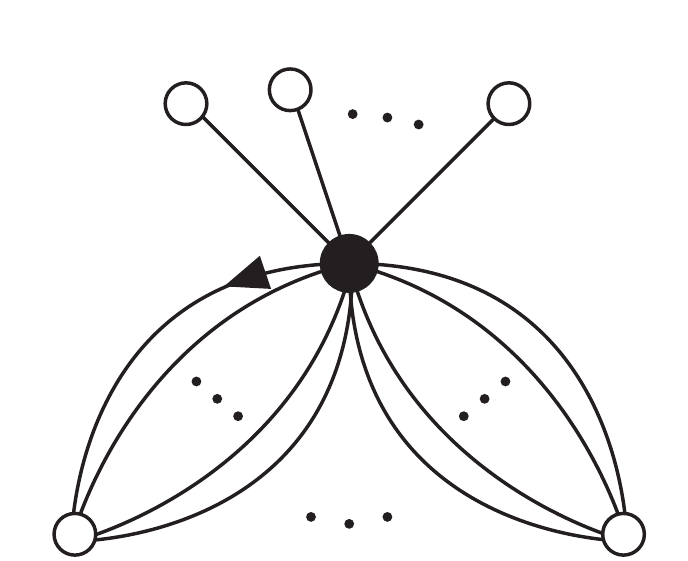}} \hspace{7mm} \mathrm{arrow\;\,not\;\,on\;\,external\;\,lines} \label{d2}
\eea}%
Here, the vertex with a black circle denotes $-R_{pq}\, C^q_{i,i_1i_2\cdots i_n}$, and the line with an arrow is the free index $p$ in the cutoff. Again, the sum is over all possible diagrams of the given form, keeping in mind that the arrow is never on the external lines. The connected $n$-point functions are still to be written in terms of the one-particle irreducible $n$-point functions, as explained in appendix \ref{feyn}, so that the diagrams on the right in \eqref{d1},\eqref{d2} will provide a representation of the left-hand side in terms of the background and the classical fluctuating fields $\bar\xi^i$. The final form of the mspWI in diagrammatic language is
\be \label{mspwi_diag} 
\boxed{\mathcal{N}_i \,\equiv\, \Gamma\!,_{i} -{\textstyle{\frac{1}{2}}}\, G R,_{i} + \;\sum\!\!\! \raisebox{-9mm}{\includegraphics[trim=7cm 22cm 7cm 2cm, clip=true, width=0.15\textwidth]{diag1}} + \;\sum\!\!\!  \raisebox{-9mm}{\includegraphics[trim=7cm 22cm 7cm 2cm, clip=true, width=0.15\textwidth]{diag2}}=0}
\ee

\subsection{Comments}

Let us pause at this point to make a few comments. In fact, eq.\eqref{mspwi_comp} can be rewritten in a different way, which also makes connection with some earlier works. This can be seen by applying eq.\eqref{O;i} to the expectation value of $\xi^n\!\!,_i$ 
\be   
\langle \xi^n\!\!,_i \rangle_{;p} = G_{pq}(\langle \xi^q \xi^n\!\!,_i \rangle - \bar\xi^q\langle \xi^n\!\!,_i \rangle),
\ee 
which, upon contraction with $G^{pm}R_{mn}$
\be  \label{id}
-G^{pm}R_{mn}\langle \xi^n\!\!,_i \rangle_{;p} = -G^{pm}R_{mn}G_{pq}(\langle \xi^q \xi^n\!\!,_i \rangle - \bar\xi^q\langle \xi^n\!\!,_i \rangle) = -\langle \xi^m R_{mn} \xi^n\!\!,_i \rangle + \bar\xi^m R_{mn} \langle \xi^n\!\!,_i \rangle,
\ee
leads exactly to the sum of the last two terms in \eqref{mspwi_comp}. So the mspWI \eqref{mspwi_comp} can also be written as
\be \label{mspwi_2}
\boxed{\Gamma\!,_{i} +\Gamma\!_{;j}\langle\xi^j\!\!,_{i} \rangle -{\textstyle{\frac{1}{2}}} G^{mn}(R_{nm}),_{i} -G^{np} R_{pm}\langle \xi^m\!\!\!,_i \rangle_{;n} =0}
\ee
In a gauge theory context, and within the geometric approach, a similar identity has been obtained in \cite{pawlowski_0310,pawlowski_review}. We emphasize that eq.\eqref{mspwi_comp} or \eqref{mspwi_2} is very general and does not rely on any assumption other than the dependence of the ultraviolet action (and the measure) on a single field. 

Before closing the discussion on the mspWI, let's also make a comment on the choice of measure in \eqref{W}. In the derivation of the mspWI, we chose the measure of integration to be a function of the total field. In the literature various choices are made for the measure, for all of which the splitting Ward identity \eqref{spwi} is claimed to be valid for the resulting effective action, see for example \cite{hps,rebhan_npb1988,burgess_kunstatter,kunstatter_1992,parker_toms}. In fact, as apparent from the steps of the derivation, any deviation from a single field measure in \eqref{W} will result in extra terms in \eqref{mspwi}, which vanish only in certain regularization schemes. For instance, the covariant measure $D\xi\sqrt{\mathrm{det}\, g_{ij}(\varphi)}$, for some metric $g_{ij}$ on field space, will lead to the mspWI 
\be \label{mspwi_5} 
\Gamma\!,_{i} +\Gamma\!_{;j}\langle\xi^j\!\!,_{i} \rangle -{\textstyle{\frac{1}{2}}}\, G^{mn}(R_{nm}),_{i} +\bar\xi^m R_{mn}\langle\xi^n\!\!,_i\rangle - \left\langle \xi^m R_{mn} \xi^n\!\!,_i\right\rangle  + \delta(0)\langle(\nabla_{\!i}\xi^j)_{;j}\rangle=0,
\ee
where $\nabla_{\!i}$ is the background covariant derivative, compatible with the metric $g_{ij}(\varphi)$, and $\delta(0)$ is the Dirac delta function in position space, evaluated at zero. 

\subsection{Consistency with the flow equation}  \label{consistency}

At any given energy scale $k$, the constraint equation \eqref{mspwi_comp}, $\mathcal{N}_{i,k}=0$, restricts the infinite dimensional theory space, i.e. the space of coefficients of all possible Lorentz (or diffeomorphism) invariant operators constructed from $\varphi^i$ and $\xi^i$, with possible internal symmetries, to a lower, but still infinite, dimensional subspace. In the ultraviolet scale the constraint identity \eqref{mspwi_comp} is automatically satisfied if the ultraviolet action $S$ depends only on the total field, which is our primary assumption, while in the infrared the identity reduces to \eqref{spwi} and tells us that the effective action is a functional of a single field $\Phi^j(\varphi,\bar\xi)$ which satisfies $\Phi^j\!,_i+\Phi^j_{;l}\langle\xi^l\!,_i\rangle=0$ (see section \ref{background} for more details). Such an interpretation doesn't seem to be possible at the intermediate scales. Roughly speaking, the dimension of these subspaces is the number of operators one can construct from a single field. 

For different values of $k$, the one-parameter set of subspaces, given by $\mathcal{N}_{i,k}=0$, sweeps a surface in theory space, of one dimension higher, on which all the curves \eqref{eaa_def} lie. In fact, the curves \eqref{eaa_def} also sweep this surface for different ultraviolet actions that depend on a single field, while satisfying the exact flow \eqref{frge}. This shows the consistency of the mspWI with the renormalization group equation \eqref{frge}. 

Of course, any solution to the exact flow equation which intersects the above mentioned surface must lie entirely on this surface by the uniqueness theorem, and in particular it must coincide with one of the trajectories \eqref{eaa_def}. In other words, starting with a solution to the mspWI at some scale, under the renormalization group flow it will remain inside the set of solutions to $\mathcal{N}_{i,k}=0$ at any other scale, and tends to some single-field dependent action in the ultraviolet\footnote{up to terms proportional to $\delta(0)$ (see eq.\eqref{kinfty}).}. 

It would still be instructive to find explicitly the flow equation for the quantity $\mathcal{N}_i$ defined in \eqref{mspwi_comp}. Some necessary relations we will be using in the following, but which in principle have a wider application, are collected in appendix \ref{flows}. By direct computation, the renormalization group flow of the first term in \eqref{mspwi_comp} is found as follows
{\setlength\arraycolsep{2pt}
\bea
\partial_t \Gamma\!,_i  &=& {\textstyle{\frac{1}{2}}} G^{mn}(\dot R_{nm}),_i - {\textstyle{\frac{1}{2}}} G^{mp}(G_{pq}),_i G^{qn}\dot R_{nm} \nn\\
&=& {\textstyle{\frac{1}{2}}} G^{mn}(\dot R_{nm}),_i - {\textstyle{\frac{1}{2}}} G^{mp}(R_{pq}),_i G^{qn}\dot R_{nm} - {\textstyle{\frac{1}{2}}} G^{mp}\Gamma_{;pq},_i G^{qn}\dot R_{nm} \nn\\
&=&  {\textstyle{\frac{1}{2}}} G^{mn}(\dot R_{nm}),_i - {\textstyle{\frac{1}{2}}} G^{qn}\dot R_{nm} G^{mp}(R_{pq}),_i  - {\textstyle{\frac{1}{2}}} G^{qn}\dot R_{nm} G^{mp}\Gamma_{;pq},_i  \nn\\
&=& {\textstyle{\frac{1}{2}}} G\dot R,_i - {\textstyle{\frac{1}{2}}} (G\dot R G)^{qp}(R_{pq}),_i  - {\textstyle{\frac{1}{2}}} (G\dot R G)^{qp}\Gamma_{;pq},_i.
\eea}%
In a similar way, the flow of the third term is found to be
{\setlength\arraycolsep{2pt}
\bea
-\frac{1}{2}\partial_t [G^{mn}(R_{nm}),_{i}] &=& -{\textstyle{\frac{1}{2}}} G^{mn}(\dot R_{nm}),_i + {\textstyle{\frac{1}{2}}} G^{mp}(\dot G_{pq}) G^{qn}(R_{nm}),_i \nn\\
&=& -{\textstyle{\frac{1}{2}}} G^{mn}(\dot R_{nm}),_i + {\textstyle{\frac{1}{2}}} G^{mp}\dot R_{pq} G^{qn}(R_{nm}),_i+ {\textstyle{\frac{1}{2}}} G^{mp}\dot \Gamma_{;pq} G^{qn}(R_{nm}),_i \nn\\
&=& -{\textstyle{\frac{1}{2}}} G\dot R,_i +{\textstyle{\frac{1}{2}}} (G\dot R G)^{mn}(R_{nm}),_i  + {\textstyle{\frac{1}{2}}} (G R,_i G)^{qp}\dot \Gamma_{;pq}.
\eea}%
Summing the two terms leads to  
\be 
\partial_t \left[\Gamma\!,_i- {\textstyle{\frac{1}{2}}} G^{mn}(R_{nm}),_{i}\right] = - {\textstyle{\frac{1}{2}}} (G\dot R G)^{qp}\Gamma_{;pq},_i  + {\textstyle{\frac{1}{2}}} (G R,_i G)^{qp}\dot \Gamma_{;pq}  = - {\textstyle{\frac{1}{2}}} (G\dot R G)^{qp}\left(\Gamma,_i\right)_{;pq}  + {\textstyle{\frac{1}{2}}} (G R,_i G)^{qp}\dot \Gamma_{;pq}.
\ee
But from \eqref{gamma_dot;ij} the last term on the right-hand side is rewritten as 
{\setlength\arraycolsep{2pt}
\bea  \label{id3}
{\textstyle{\frac{1}{2}}} (G R,_i G)^{qp}\dot \Gamma_{;pq} 
&=& - {\textstyle{\frac{1}{4}}} (G R,_i G)^{qp} (G\dot R G)^{rs} \left[\Gamma_{;pqrs} + 2 G^{mn} \Gamma_{;prm} \Gamma_{;qsn}\right]  \nn\\
&=& - {\textstyle{\frac{1}{4}}} (G \dot R G)^{qp} (G R,_i G)^{rs} \left[\Gamma_{;pqrs} + 2 G^{mn} \Gamma_{;prm} \Gamma_{;qsn}\right]  = - {\textstyle{\frac{1}{2}}} (G\dot R G)^{qp}\left[-{\textstyle{\frac{1}{2}}}GR,_{i}\right]_{;pq},
\eea}%
where in the second equation we have used the fact that the term inside the brackets is symmetric with respect to the interchange $qp\leftrightarrow rs$, and therefore we have interchanged $\dot R$ and $R,_i$, and in the final equation we have used an identity similar to \eqref{gamma_dot;ij} but with $\dot R$ replaced by $R,_i$. So finally we find the equation
\be 
\partial_t \left[\Gamma\!,_i-{\textstyle{\frac{1}{2}}} G^{mn}(R_{nm}),_{i}\right] = - {\textstyle{\frac{1}{2}}} (G\dot R G)^{qp}\left[\Gamma\!,_i-{\textstyle{\frac{1}{2}}} G^{mn}(R_{nm}),_{i}\right]_{;pq},
\ee
which relates the scale derivative of the sum of the first and third terms in \eqref{mspwi_comp} to the second $\xi$-derivatives of the same quantity. Now let us consider the remaining terms in \eqref{mspwi_comp}. For the second term we find a flow of the following form
{\setlength\arraycolsep{2pt}
\bea
\partial_t [\Gamma\!_{;j}\langle\xi^j\!\!,_{i} \rangle] 
&=& \partial_t \Gamma\!_{;j}\langle\xi^j\!\!,_{i} \rangle + \Gamma\!_{;j}\partial_t\langle\xi^j\!\!,_{i} \rangle \nn\\
&=&  - {\textstyle{\frac{1}{2}}} (G\dot R G)^{qp}(\Gamma\!_{;j})_{;pq}\, \langle\xi^j\!\!,_{i} \rangle - {\textstyle{\frac{1}{2}}}\Gamma\!_{;j}\, (G\dot R G)^{qp} \langle\xi^j\!\!,_{i} \rangle_{;pq} \nn\\
&=& - {\textstyle{\frac{1}{2}}} (G\dot R G)^{qp}(\Gamma\!_{;j}\langle\xi^j\!\!,_{i} \rangle)_{;pq} + (G\dot R G)^{qp}\Gamma\!_{;jp}\, \langle\xi^j\!\!,_{i} \rangle_{;q} \nn\\
&=& - {\textstyle{\frac{1}{2}}} (G\dot R G)^{qp}(\Gamma\!_{;j}\langle\xi^j\!\!,_{i} \rangle)_{;pq} + (G\dot R G)^{qp}G_{jp}\, \langle\xi^j\!\!,_{i} \rangle_{;q} -  (G\dot R G)^{qp}R_{jp}\, \langle\xi^j\!\!,_{i} \rangle_{;q} \nn\\
&=& - {\textstyle{\frac{1}{2}}} (G\dot R G)^{qp}(\Gamma\!_{;j}\langle\xi^j\!\!,_{i} \rangle)_{;pq} + (G\dot R)^q_j\, \langle\xi^j\!\!,_{i} \rangle_{;q} -  (G\dot R G R)^q_j\, \langle\xi^j\!\!,_{i} \rangle_{;q} 
\eea}%
while, using \eqref{Odot} we can write the $t$-derivative of the last two terms in \eqref{mspwi_comp} as
{\setlength\arraycolsep{2pt}
\bea
\partial_t (\bar\xi^m R_{mn}\langle\xi^n\!\!,_i\rangle - \left\langle \xi^m R_{mn} \xi^n\!\!,_i\right\rangle) 
&=&  - {\textstyle{\frac{1}{2}}} (G\dot R G)^{qp}\bar\xi^m R_{mn}\langle\xi^n\!\!,_i\rangle_{;pq} + \bar\xi^m \dot R_{mn}\langle\xi^n\!\!,_i\rangle \nn\\
&& - {\textstyle{\frac{1}{2}}} (G\dot R G)^{qp}(- \left\langle \xi^m R_{mn} \xi^n\!\!,_i\right\rangle)_{;pq} - \langle \xi^m \dot R_{mn} \xi^n\!\!,_i\rangle \nn\\[2mm]
&=& - {\textstyle{\frac{1}{2}}} (G\dot R G)^{qp}(\bar\xi^m R_{mn}\langle\xi^n\!\!,_i\rangle)_{;pq} +(G\dot R G)^{qp}R_{pn}\langle\xi^n\!\!,_i\rangle_{;q} + \bar\xi^m \dot R_{mn}\langle\xi^n\!\!,_i\rangle \nn\\
&& -{\textstyle{\frac{1}{2}}} (G\dot R G)^{qp}(- \left\langle \xi^m R_{mn} \xi^n\!\!,_i\right\rangle)_{;pq} - \langle \xi^m \dot R_{mn} \xi^n\!\!,_i\rangle \nn\\[2mm]
&=& -{\textstyle{\frac{1}{2}}} (G\dot R G)^{qp}(\bar\xi^m R_{mn}\langle\xi^n\!\!,_i\rangle -\langle \xi^m R_{mn} \xi^n\!\!,_i\rangle)_{;pq} +(G\dot R G R)^q_n\langle\xi^n\!\!,_i\rangle_{;q} \nn\\
&& + \bar\xi^m \dot R_{mn}\langle\xi^n\!\!,_i\rangle  - \langle \xi^m \dot R_{mn} \xi^n\!\!,_i\rangle. 
\eea}%
Again, summing the two pieces gives
{\setlength\arraycolsep{2pt}
\bea
\partial_t [\Gamma\!_{;j}\langle\xi^j\!\!,_{i}\rangle + \bar\xi^m R_{mn}\langle\xi^n\!\!,_i\rangle -  \langle\xi^m R_{mn} \xi^n\!\!,i \rangle] &=& - {\textstyle{\frac{1}{2}}} (G\dot R G)^{qp}[\Gamma\!_{;j}\langle\xi^j\!\!,_{i}\rangle + \bar\xi^m R_{mn}\langle\xi^n\!\!,_i\rangle -\langle \xi^m R_{mn} \xi^n\!\!,_i\rangle]_{;pq} \nn\\
&& + (G\dot R)^q_j\, \langle\xi^j\!\!,_{i} \rangle_{;q}  + \bar\xi^m \dot R_{mn}\langle\xi^n\!\!,_i\rangle  - \langle \xi^m \dot R_{mn} \xi^n\!\!,_i\rangle.
\eea}%
The second line vanishes by an identity similar to \eqref{id}, which is found by applying \eqref{O;i} to $O =\xi^j\!\!,_{i}$, and contracting with $(G\dot R)^q_j$. Thus a relation similar to that of the sum of the first and third terms in \eqref{mspwi_comp} holds also for the sum of the remaining terms. So we have shown that for the quantities $\mathcal{N}_{1i}$ and $\mathcal{N}_{2i}$ defined \nolinebreak by
\be \label{N_12}
\mathcal{N}_i \equiv \underbrace{\Gamma\!,_{i} -{\textstyle{\frac{1}{2}}}\, G^{mn}(R_{nm}),_{i}}_{\mathcal{N}_{1i}} +\underbrace{\Gamma\!_{;j}\langle\xi^j\!\!,_{i} \rangle +\bar\xi^m R_{mn}\langle\xi^n\!\!,_i\rangle - \left\langle \xi^m R_{mn} \xi^n\!\!,_i\right\rangle}_{\mathcal{N}_{2i}} 
\ee
we have
\be \label{N12}
\partial_t \mathcal{N}_{1i} =  - {\textstyle{\frac{1}{2}}}\, (G\dot R G)^{qp}(\mathcal{N}_{1i})_{;pq}, \hspace{5mm} \partial_t \mathcal{N}_{2i} =  - {\textstyle{\frac{1}{2}}}\, (G\dot R G)^{qp}(\mathcal{N}_{2i})_{;pq},
\ee
and consequently the same identity is valid for their sum $\mathcal{N}_i$
\be \label{N}
\boxed{\partial_t \mathcal{N}_i =  - {\textstyle{\frac{1}{2}}}\, (G\dot R G)^{qp}(\mathcal{N}_i)_{;pq}}
\ee
This gives an explicit equation for the running of the quantity $\mathcal{N}_i$ with the energy scale, and fulfills our expectations elaborated on at the beginning of this section. The extra information \eqref{N12} will be used in section \ref{gauge} to address background gauge invariance of the geometric effective action for gauge theories. Equation \eqref{N} is also similar to a flow equation, reported in \cite{pawlowski_0310} for gauge theories in the geometric approach.

\section{Background dependence} \label{background}

\subsection{General considerations}

At the beginning of section \ref{consistency} we briefly pointed out that in the infrared limit $k=0$, as a result of the splitting Ward Identity \eqref{spwi}, the effective action depends on a single field $\Phi^l$ which is implicitly defined through $\Phi^l\!,_{i} +\Phi^l\!_{;j}\langle\xi^j\!\!,_{i} \rangle =0$\footnote{Here we are assuming that this differential equation is integrable, with a solution $\Phi(\varphi,\xi)$ that is invertible as a function of $\xi$. The case of linear splitting, studied in the next subsection, provides the simplest but not the only such example. The following argument is therefore valid only if the this assumption is \nolinebreak true.}. This suggests re-expressing the effective action in terms of the background and the field $\Phi^i$ by defining the quantity $\bar\Gamma_{\!k}[\varphi,\Phi]$ at an arbitrary scale as
\be \label{bargamma}
\Gamma_{\!k}[\varphi,\xi] = \bar\Gamma_{\!k}[\varphi,\Phi], \hspace{1cm}
\Phi^l\!,_{i} +\Phi^l\!_{;j}\langle\xi^j\!\!,_{i} \rangle =0,
\ee
where the field $\Phi_k(\varphi,\xi)$ is considered as a function of $\varphi^i$, $\xi^i$ and $k$. Let us now rewrite the mspWI in terms of $\bar\Gamma_k$. The partial derivatives of the two effective actions are related in the following way
{\setlength\arraycolsep{2pt}
\bea     
\frac{\delta\Gamma_{\!k}[\varphi,\xi]}{\delta\varphi^i} &=& \frac{\delta\bar\Gamma_{\!k}[\varphi,\Phi]}{\delta\varphi^i} + \frac{\delta\bar\Gamma_{\!k}[\varphi,\phi]}{\delta\Phi}\!\cdot\!\frac{\delta\Phi}{\delta\varphi^i}, \label{barvarphi} \\
\frac{\delta\Gamma_{\!k}[\varphi,\xi]}{\delta\xi^i} &=& \frac{\delta\bar\Gamma_{\!k}[\varphi,\Phi]}{\delta\Phi}\!\cdot\!\frac{\delta\Phi}{\delta\xi^i}. \label{barxi}
\eea}%
Using these relations, and the definition of $\Phi^i$, in \eqref{mspwi_2} gives the background dependence of $\bar\Gamma_{\!k}[\varphi,\Phi]$ 
\be \label{mspwi_3}
\bar \Gamma\!,_{i} = {\textstyle{\frac{1}{2}}} G^{mn}(R_{nm}),_{i}+ G^{np} R_{pm}\langle \xi^m\!\!\!,_i \rangle_{;n} .
\ee
It is clear that in the infrared, $k=0$, all the background dependence is gone, and the effective action is a function of the single field $\Phi^i$, as was expected by construction.

If we further \textit{assume} that the scale and background dependence of the cutoff kernel can be collected into a dependence on a single quantity $\hat k(k,\varphi)$ \cite{morris_1312}, then this is even more simplified. In this case we define 
\be \label{hatgamma}
\Gamma_{\!k}[\varphi,\xi] = \hat\Gamma_{\!\hat k}[\varphi,\Phi], \hspace{1cm}
\Phi^l\!,_{i} +\Phi^l\!_{;j}\langle\xi^j\!\!,_{i} \rangle =0, \hspace{5mm}
R_k(\varphi) = \hat R_{\hat k}, 
\ee
where, here, the field $\Phi_{\hat k}(\varphi,\xi)$ is considered as a function of $\varphi^i$, $\xi^i$ and $\hat k$, and the partial derivatives in its defining equation above are defined accordingly. Notice also that with the assumption $R_k(\varphi) = \hat R_{\hat k}$ the quantity $\langle\xi^j\!\!,_{i} \rangle$ can be considered as a function of $\varphi^i$, $\xi^i$ and $\hat k$. Compared to the previous situation, the relation between partial derivatives of the effective actions is modified due to the field dependence of $\hat k$
{\setlength\arraycolsep{2pt}
\bea     
\frac{\delta\Gamma_{\!k}[\varphi,\xi]}{\delta\varphi^i} &=& \frac{\delta\hat\Gamma_{\!\hat k}[\varphi,\Phi]}{\delta\varphi^i} + \frac{\delta\hat\Gamma_{\!\hat k}[\varphi,\Phi]}{\delta\Phi}\!\cdot\!\frac{\delta\Phi}{\delta\varphi^i} + \frac{\delta\hat\Gamma_{\!\hat k}[\varphi,\Phi]}{\delta\Phi}\!\cdot\!\frac{\delta\Phi}{\delta\hat k}\frac{\delta\hat k}{\delta\varphi^i} + \frac{\delta\hat\Gamma_{\!\hat k}[\varphi,\Phi]}{\delta\hat k}\frac{\delta\hat k}{\delta\varphi^i}, \label{hatvarphi} \\
\frac{\delta\Gamma_{\!k}[\varphi,\xi]}{\delta\xi^i} &=& \frac{\delta\hat\Gamma_{\!\hat k}[\varphi,\Phi]}{\delta\Phi}\!\cdot\!\frac{\delta\Phi}{\delta\xi^i}. \label{hatxi}
\eea}%
In terms of $\hat \Gamma_{\!\hat k}$ the flow equation is also modified. Dropping the term $\partial\hat t/\partial t|_{\varphi,\xi}$ ($\hat t = \log \hat k$), in which $\varphi^i$ and $\xi^i$ are held fixed, the modified flow equation will be
\be \label{hatt}
\frac{\partial\hat\Gamma_{\!\hat k}[\varphi,\Phi]}{\partial\hat t} + \frac{\delta\hat\Gamma_{\!\hat k}[\varphi,\Phi]}{\delta\Phi}\!\cdot\!\frac{\partial\Phi}{\partial\hat t} = \frac{1}{2} G \frac{\partial \!\hat R_{\hat k}}{\partial\hat t}.
\ee
Compared to \eqref{barvarphi}, there are two extra terms in \eqref{hatvarphi}, which, using the above relation, cancel the first term on the right-hand side of \eqref{mspwi_3}. The mspWI then reduces to the simple identity
\be \label{mspwi_4}
\hat \Gamma\!,_{i} = G^{np} R_{pm} \langle \xi^m\!\!\!,_i \rangle_{;n}.
\ee
One can also easily check that moving to dimensionless variables $\Phi^i = \hat k^D \hat \Phi^i$ denoted by a hat, where $D$ is the dimension of the field $\Phi^i$, and defining $\hat\Gamma_{\!\hat k}[\varphi,\Phi]=\widehat \Gamma_{\!\hat k}[\varphi,\hat\Phi]$, equations \eqref{hatt} and \eqref{mspwi_4} are still valid with $\hat\Gamma$ replaced by $\widehat{\Gamma}$, and $\Phi^i$ replaced by $\hat\Phi^i$. In general, the function $\hat k$ can be read off from the condition $R_k(\varphi) = \hat R_{\hat k}$, if valid, and the redefinitions of the dynamical field and the action are found from \eqref{bargamma} or \eqref{hatgamma}. It is worth emphasizing that, as evident from \eqref{mspwi_3} and \eqref{mspwi_4}, in the limit $k\rightarrow 0$, background independence of $\bar\Gamma$ and $\hat\Gamma$ will be restored regardless of how the total field is split.

\subsection{A special case} 

Given the general analysis above, it is now straightforward to reproduce some results in the literature. In the special case of linear splitting the right-hand side of \eqref{mspwi_4} vanishes because $\xi^j\!\!,_i = -\delta^j_i$ and therefore $\langle \xi^m\!\!,_i \rangle_{;n}=0$. So in such examples where $R_k(\varphi) = \hat R_{\hat k}$ and where the fields are split linearly, complete background independence of the effective action $\hat\Gamma_{\!\hat k}$ or $\widehat\Gamma_{\!\hat k}$ is guaranteed by the identity \eqref{mspwi_4}. Also the required field redefinition follows trivially from the middle equation in \eqref{hatgamma}, which reduces to $\Phi^l\!,_{i} -\Phi^l\!_{;i}=0$, and suggests $\Phi^i = \varphi^i + \xi^i \equiv \phi^i$. In this case, equations \eqref{hatt} and \eqref{mspwi_4} reduce to 
\be 
\partial_{\hat t}\hat\Gamma = {\textstyle{\frac{1}{2}}} G \partial_{\hat t} \hat R_{\hat k}, \qquad 
\hat \Gamma\!,_{i} = 0,
\ee
while for $\widehat\Gamma_{\!\hat k}$ they simplify to the following equations
\be 
\partial_{\hat t}\widehat\Gamma - D \hat\Phi^i \widehat\Gamma_{;i} = {\textstyle{\frac{1}{2}}} G \partial_{\hat t} \hat R_{\hat k}, \qquad 
\widehat \Gamma\!,_{i} = 0.
\ee
Examples in the literature where the assumption $R_k(\varphi) = \hat R_{\hat k}$ is valid and where the linear splitting is performed are discussed in the context of scalar field theory for a special kind of cutoff \cite{morris_1312}, and also conformally reduced gravity \cite{morris_1502}, for which, in two spacetime dimensions and when there is no anomalous dimension for the field $\xi^i$, the assumption is valid for any cutoff just based on dimensional grounds. 

\section{Modified master equation}

\subsection{Motivation and derivation of the identity}

The mspWI \eqref{mspwi_comp},\eqref{mspwi_2},\eqref{mspwi_diag} is supposed to put constraints on the form of the effective action, which would otherwise be a general functional of the background field and the fluctuations, compatible with other possible imposed symmetries. In practice, there are, however, two obstacles before using the mspWI to constrain the effective action. 

First of all, equation \eqref{mspwi_diag} is actually divergent, because the loop diagrams in the third term, and the diagrams in the fourth term which have loops without an arrow, introduce infinities. It is therefore not possible to use this equation directly to put constraints on the renormalized effective action, which is what we finally insert into the exact flow equation. In order to remove the divergences we need to perform a loop expansion to the desired accuracy and introduce counter-terms order by order in perturbation theory. 
But in this case, there is no point in using the mspWI, because the renormalization group equation \eqref{frge} itself already gives the flow of the effective action at any loop order \cite{litim_0111,codello_1310,codello_1312}, which can be solved iteratively to get the $l$-loop effective action, and the mspWI is automatically satisfied by these solutions at any order of perturbation (see subsection \ref{loop_exp}). On the other hand, if we are not interested in a loop expansion but instead willing to perform another kind of approximation, e.g. an expansion in the number of derivatives and the order of the fluctuating fields, then we will face, once again, the problem of  divergences in \eqref{mspwi_diag}. 

Second, in such a situation, where, instead of doing perturbation theory in the number of loops, we are interested, say, in a derivative expansion and an expansion in the number of fluctuations, then from \eqref{mspwi_diag} it is seen that at each level (order of fluctuations) we need to take into account an infinite number of diagrams, of the type of the third and fourth terms, which include diagrams of all possible loop numbers. This is, of course, practically impossible.
Both of these problems could be overcome if we were able to write the constraint equation \eqref{mspwi_comp} entirely in terms of the effective action and its derivatives, with no manifest divergent loop terms. This is achieved by following the BRS idea. 

According to the BRS prescription, the action is modified by introducing a source term $I_j$ for the variation of the quantum field, $c^i \xi^j\!\!,_i$. In order for this new action to be invariant under the transformations \eqref{symm}, the transformation parameter $c^i$ is taken to be a Grassmannian variable, and the source, which is consequently forced to be a Grassmannian field, must be itself invariant under the symmetry transformations. In sum, the infinitesimal symmetry transformation operator for a general function of these fields is 
\be  \label{s}
s \equiv c^i \frac{\delta}{\delta\varphi^i} + c^i \xi^j\!\!,_i \frac{\delta}{\delta\xi^j},
\ee
where, in general, the partial derivatives are taken, with other fields, from the set $\varphi^i, \xi^i, c^i$ and $I_i$, kept fixed. In particular, the fields themselves transform in the following way:
\be \label{sfield}
s \varphi^i = c^i, \hspace{5mm} s \xi^i = c^j \xi^i\!\!,_j, \hspace{5mm} s c^i =0, \hspace{5mm} s I_i =0.
\ee
The generator of connected diagrams is now also a function of the source field $I_i$ and the transformation variable $c^i$. Explicitly\footnote{The same symbols $W$ and $\Gamma$ , as their $I_i=0$ counterparts are used in this section to avoid complicating the notation.}
\be  \label{WK}
\exp\left(-W_k[\varphi,c,I,J]\right) 
= \int \!\! D\phi \;\mu(\phi) \;\, \mathrm{exp}\left(-\Sigma[\varphi,\xi,c,I]-S_k[\varphi,\xi]- J_i\xi^i\right), \hspace{3mm} \Sigma[\varphi,\xi,c,I] \equiv S[\phi]+I_i s\xi^i. 
\ee
The modified action $\Sigma$ is invariant under the infinitesimal transformations \eqref{sfield}. The corresponding symmetry constraint for the effective action $\Gamma[\varphi,\bar\xi,c,I]$ follows trivially along the lines of the proof of \eqref{mspwi_2} and is very similar to this equation, except for the appearance of the Grassmannian transformation variable $c^i$
\be 
c^i\Gamma\!,_{i} -{\textstyle{\frac{1}{2}}}\, G\,s R +\Gamma\!_{;j}\langle s\xi^j \rangle -G^{np}R_{pm}\langle s\xi^m \rangle_{;n} =0
\ee
The Ward identity we have been looking for, now follows immediately by noticing that 
\be 
\langle s\xi^i \rangle = \frac{\delta W}{\delta I_i} = \frac{\delta\Gamma}{\delta I_i} \equiv \Gamma^i.
\ee
Denoting, for conciseness, a partial derivative with respect to the sources $I_i$, with an upper index as above, the result will be
\be \label{master}
\boxed{c^i\Gamma\!,_{i} +\Gamma^j\Gamma\!_{;j} -{\textstyle{\frac{1}{2}}}\, G^{mn}\,s R_{nm} -G^{np}R_{pm}\Gamma^m_{;n} =0}
\ee
This is the desired Ward identity, or \textit{modified master equation}. This gives an equation written entirely in terms of the effective action and its derivatives but with no manifest divergent loop terms. In other words, if the effective action were finite, there would be no divergent terms in this equation. There are actually two terms, the third and fourth ones in \eqref{master}, with a manifest loop. These are, however, regulated with $R_k$, and therefore introduce no divergences, when computed with a finite effective action.

\subsection{Loop expansion} \label{loop_exp}

It would be instructive to see explicitly how the modified master equation \eqref{master} is satisfied at tree level and especially at one-loop level. Notice that \eqref{master} reduces to \eqref{mspwi_comp} upon setting $I_i=0$, and therefore provides a generalization to that. To begin with, let's write the tree level master equation. The last two terms in \eqref{master} already have a loop. So the tree level part of this equation is 
\be \label{master0}
c^i\Sigma,_{i} +\Sigma^j\Sigma_{;j}  =0.
\ee
Using \eqref{s}, this can also equivalently be written as $s\Sigma =0$, which is trivially satisfied by construction. The one-loop term is also easily found to be
\be \label{master1_1}
D_\Sigma\Gamma^{1-\mathrm{loop}} -{\textstyle{\frac{1}{2}}}\, G_0^{mn}\,s R_{nm} -G_0^{np}R_{pm}\Sigma^m_{;n} =0,
\ee
where we have defined the nilpotent differential operator
\be \label{dsigma}
D_\Sigma \equiv c^i \frac{\delta}{\delta\varphi^i} + \Sigma_{;j} \frac{\delta}{\delta I_j} + \Sigma^j \frac{\delta}{\delta\xi^j},
\ee
and by $G^{mn}_0$ we mean the propagator, in which the tree level action $\Sigma$ has been used instead of the effective action. After a bit of manipulation, \eqref{master1_1} can be brought into the following form
\be \label{master1_2}
D_\Sigma\left[\Gamma^{1-\mathrm{loop}} +{\textstyle{\frac{1}{2}}}\mathrm{Tr} \log(1-G_0R)\right] =0.
\ee
This is nothing but the one-loop (unmodified) master equation. In fact the second term in the brackets is the difference between the one-loop effective action with a regulator and the one without a regulator, $\left.\Gamma^{(1)}\right|_{R=0} - \Gamma^{(1)}$, so that the quantity inside the brackets will be the one-loop effective action in the absence of a regulator. Consequently, in order to verify eq.\eqref{master1_1} or \eqref{master1_2}, we need to check if the usual one-loop effective action in the absence of a regulator, 
\be \label{trlog}
{\textstyle{\frac{1}{2}}}\mathrm{Tr} \log \Sigma^{(2)},
\ee
vanishes under the action of $D_\Sigma$. Using the definition \eqref{dsigma}, and denoting the matrix of second $\xi$-derivatives of $\Sigma$ by $\Sigma_{(2)}$, and its inverse by $\Sigma^{(2)}$, we get 
\be  \label{dtrlog}
D_\Sigma\, {\textstyle{\frac{1}{2}}}\mathrm{Tr} \log \Sigma^{(2)} = {\textstyle{\frac{1}{2}}}\mathrm{Tr} \;\Sigma_{(2)}\, c^i\Sigma,_i^{(2)} +{\textstyle{\frac{1}{2}}}\mathrm{Tr}\; \Sigma_{(2)}\, \Sigma_{;j}\Sigma^{(2)j} + {\textstyle{\frac{1}{2}}}\mathrm{Tr}\; \Sigma_{(2)}\, s\xi^j\Sigma_{;j}^{(2)}, \hspace{5mm} \Sigma_{(2)}\Sigma^{(2)} =1.
\ee
In order to simplify this further, one can find a relation between derivatives of the tree level action by taking the second $\xi$-derivative of the tree level equation \eqref{master0}
\be 
c^i\Sigma\!_{;mn,i} +(s\xi^j)_{;mn}\Sigma_{;j}+s\xi^j\Sigma_{;jmn}  = -(s\xi^j)_{;m}\Sigma_{;jn} -(s\xi^j)_{;n}\Sigma_{;jm}.
\ee
Inserting this into \eqref{dtrlog} leads to 
\be 
D_\Sigma\, {\textstyle{\frac{1}{2}}}\mathrm{Tr} \log \Sigma^{(2)} 
= -{\textstyle{\frac{1}{2}}} \;\Sigma^{nm}_{(2)}\, (s\xi^j)_{;m}\Sigma^{(2)}_{;jn} -{\textstyle{\frac{1}{2}}}\; \Sigma^{mn}_{(2)}\, (s\xi^j)_{;n}\Sigma^{(2)}_{;jm} = -\delta(0)\; (s\xi^j)_{;j}  \label{dstrlog}
\ee
This shows that \eqref{trlog} doesn't generally vanish under the action of $D_\Sigma$. There is, in fact, a contribution to the one-loop effective action that we have been missing, which comes from the path integral measure. Indeed, the one-loop effective action is given by \eqref{trlog} only for the measure $D\xi$, so we need to take into account the extra terms in $D\phi\,\mu[\phi]$. The factor $\mu(\phi)$ in front of $D\phi$, which is a function of the total field can be exponentiated and counts as one-loop as it has no factor of $\hbar$. Clearly, this one-loop term vanishes under the action of $D_\Sigma$, because it is a function of the total field , and does not depend on the source $I_i$. Still, changing variables from $\phi^i$ to $\xi^i$ introduces a Jacobian 
\be 
D\phi 
= D\xi \;\,\mathrm{det}\!\left.\frac{\delta\phi^i}{\delta\xi^j}\right|_{\varphi} 
= D\xi\;\,\exp\left(-\log \mathrm{det}\!\left.\frac{\delta\xi^i}{\delta\phi^j}\right|_{\varphi}\right) 
=D\xi\;\,\exp\left(-\mathrm{Tr}\log \left.\frac{\delta\xi^i}{\delta\phi^j}\right|_{\varphi}\right),
\ee
that contributes to the effective action at one-loop. Therefore, ignoring the term $-\log\mu[\phi]$, we have 
\be 
\Gamma^{1-\mathrm{loop}} = \mathrm{Tr}\log \left.\frac{\delta\xi^i}{\delta\phi^j}\right|_{\varphi} + {\textstyle{\frac{1}{2}}}\mathrm{Tr} \log \Sigma^{(2)}.
\ee
Notice that the trace is taken with respect to the generalized indices including the spacetime points in $\xi^i$ and $\phi^i$, so the first term is actually proportional to $\delta(0)$. Let's now see what the action of $D_\Sigma$ is on this new term
{\setlength\arraycolsep{2pt}
\bea
D_\Sigma \,\mathrm{Tr}\log  \left.\frac{\delta\xi^i}{\delta\phi^j}\right|_{\varphi}
&=& \mathrm{Tr} \left(\left.\frac{\delta\phi}{\delta\xi}\right|_{\varphi}\! D_\Sigma\left.\frac{\delta\xi}{\delta\phi}\right|_{\varphi}\right)
= \left.\frac{\delta\phi^k}{\delta\xi^j}\right|_{\varphi}\! \left. c^i\frac{\delta}{\delta\varphi^i}\right|_{\phi}\left.\frac{\delta\xi^j}{\delta\phi^k}\right|_{\varphi} = \left.\frac{\delta}{\delta\xi^j}\right|_{\varphi}c^i\left.\frac{\delta\xi^j}{\delta\varphi^i}\right|_{\phi} = \delta(0)\; (s\xi^j)_{;j}.
\eea}%
To get the third equation, we have changed the order of $\varphi^i$ and $\phi^k$ differentiations and contracted the $k$ index. This cancels \eqref{dstrlog} exactly.

\subsection{Renormalization}

The master equation is normally used to prove renormalizability (if there) at least in its modern sense of providing an algorithm to remove ultraviolet divergences order by order in a loop expansion by an appropriate choice of parameters in the bare theory. This may require employing specific regularization schemes. We restrict to theories renormalizable in this sense. In particular we assume the stability of \eqref{master0}, i.e. that counter-terms can be introduced in such a way that the structure of the tree level identity \eqref{master0} is maintained for the renormalized action $\Sigma_r = \Sigma -$ counter-terms. The renormalization program can be carried out in the same way also in the presence of the infrared regulator. Since the difference $\Gamma_k - \Gamma_0$ is a finite quantity, the counter-terms required to render $\Gamma_k$ finite are the same as those of $\Gamma_0$, and satisfy
\be
D_\Sigma \,\Gamma^{l-\mathrm{loop}}_{\mathrm{div}} =0,
\ee
at the $l$-loop order, as can be seen from \eqref{master}. The modified master equation therefore provides no further information in this respect. However, removing the divergences by adding counter-terms to the ultraviolet action, we end up with the modified master equation for the renormalized ($I_i$-dependent) effective action $\Gamma_r$
\be \label{master_ren}
c^i\Gamma\!_{r,i} +\Gamma_r^j\Gamma\!_{r;j} -{\textstyle{\frac{1}{2}}}\, G_r^{mn}\,s R_{nm} -G_r^{np}R_{pm}\Gamma^m_{r;n} =0,
\ee
where $G^{mn}_r$ is the inverse of $(G_r)_{mn} = (\Gamma_r)_{;mn} + R_{mn}$. This  equation is finite and thus can be used to put constraints on the form of the renormalized effective action, at arbitrary energy scales. There is however a price to pay, and that is that 
one needs to take into account the dependence on the extra source field $I_i$ as well, when writing the most general ansatz. The field $I_i$ can finally be set to zero in \eqref{master_ren}, in which case $\Gamma_r$ will be the renormalized effective action.

The problems pointed out at the beginning of this section are generically also encountered in the definition of the field $\Phi^i$ in \eqref{bargamma},\eqref{hatgamma}. These can be similarly overcome by replacing $\langle\xi^j\!\!,_i\rangle$ with $\Gamma^j_{r;i}$ in \eqref{bargamma},\eqref{hatgamma}, which gives the renormalized $\langle\xi^j\!\!,_i\rangle$ when evaluated at $I_i=0$. The resulting equations provide a definition for $\Phi_r$, which is then to be used along with \eqref{master_ren}, leading to similar results as \eqref{mspwi_3},\eqref{mspwi_4}, with $\langle\xi^j\!\!,_i\rangle$ replaced by $\Gamma^j_{r;i}$. Finally, let us note that the flows \eqref{N12} are also valid for both \eqref{master} and \eqref{master_ren}.

\section{Covariant effective action} \label{cef}

Let us emphasize again that the results we have obtained so far are general in the sense that they do not depend on how the total field is split. However, for a general field splitting, the effective action defined using \eqref{W} and \eqref{eaa_def} is not in general covariant, that is, for a field transformation $\phi^i\rightarrow\phi'^i$ (and accordingly $\varphi^i\rightarrow\varphi'^i$, $\xi^i\rightarrow\xi'^i$), starting with the transformed action $S'$ which satisfies $S'[\phi']=S[\phi]$, and the transformed measure $\mu\rightarrow\mu'$, in the path integral, will not necessarily lead to an effective action $\Gamma'$ for which $\Gamma'[\varphi',\bar\xi']=\Gamma[\varphi,\bar\xi]$. In other words, the effective action is not a scalar under field redefinitions prior to quantization.  

As first demonstrated by Vilkovisky \cite{vilkovisky_gospel1984, vilkovisky_npb1984} and DeWitt \cite{dewitt_effact}, in order to have a covariant effective action, the quantum fields must be defined such as to transform as vectors of the field space, and moreover, the measure must be reparametrization invariant, in the sense $D\phi'\, \mu'[\phi'] = D\phi\, \mu[\phi]$, or more generally $D\phi'\; \mu'[\phi']\exp(-S'[\phi']) = D\phi\; \mu[\phi] \exp(-S[\phi])$.

According to the methods developed in \cite{honerkamp_npb1972,alvarez_annp1981,boulware_annp1982}, a natural way to achieve a vector dynamical field is to use the exponential parametrization, where the total field is given by the action of the exponential map on the fluctuations at the base point of the background field, $\phi = Exp_\varphi \xi$. For this purpose the field space must be equipped with a connection $\Gamma^k_{ij}$. One can use the connection to define a geodesic curve $\gamma^i$, in the affine parametrization, as the solution to $\ddot\gamma^k+\Gamma^k_{ij}\dot\gamma^i\dot\gamma^j=0$, with a dot on $\gamma^i$ indicating a derivative with respect to its argument. The exponential function $Exp_\varphi$ at the background point $\varphi^i$ is then defined to map a vector $\xi^i$ at $\varphi^i$ to a point $\gamma^i(1)\equiv \phi^i$ given by the geodesic evaluated at unit value of its argument, where the geodesic passes through $\varphi^i= \gamma^i(0)$, tangent to $\xi^i= \dot\gamma^i(0)$, at zero value of its argument. Some explicit results on the exponential parametrization are collected in appendix \ref{exp}. 


It will be more economic to have a metric $g_{ij}$ on field space. This can be used to define the connection and furthermore a covariant measure $D\phi \,\sqrt{\det g_{ij}(\phi)}$. The generator of connected diagrams is now given by
\be  \label{W_cov}
\exp\left(-W_k[\varphi,J]\right) = \int \!\! D\phi \,\sqrt{\det g_{ij}(\phi)} \; \mathrm{exp}\left(-S[\phi]-S_k[\varphi,\xi]- J\!\cdot\!\xi\right).
\ee
This is also covariant in the sense $W'_k[\varphi',J'] = W_k[\varphi,J]$, where $J_i$ transforms as a (lower index) covariant vector, and $W'$ is defined with the transformed metric $g'$ in the measure, and the transformed action $S'$ in the exponent. 

Using the covariant formulation, the mspWI is expected to take the same form in any coordinate system. This is, however, not manifest in \eqref{mspwi_comp} or \eqref{mspwi_2}, particularly because of the presence of ordinary background derivatives of the vector $\xi^i$ and the cutoff $R_{ij}$ in these equations. But a closer look reveals that the first term $\Gamma\!,_i$ is not covariant either. This is due to the fact that the ordinary background derivative is taken while keeping the vector $\xi^i$ fixed. This derivative, although legitimate, does not have a geometrical interpretation because the vector $\xi^i$ is defined at the base point of the background field $\varphi^i$. One can therefore write the effective action in a more useful way by expressing it in terms of the background and the total field with a bar $\bar\phi\equiv Exp_\varphi\bar\xi$, in which case we use a tilde on the effective action $\Gamma[\varphi,\bar\xi(\varphi,\bar\phi)]= \tilde \Gamma[\varphi,\bar\phi(\varphi,\bar\xi)]$. Note that the total field $\bar\phi$ shouldn't be confused with the expectation value $\langle\phi\rangle = \langle Exp_\varphi\xi\rangle$.

The background derivative of the effective action keeping the total field fixed $\tilde \Gamma\!,_i$ is now a covariant vector, and can be written as 
\be \label{gammatilde}
\tilde \Gamma\!,_i = \Gamma\!,_i + \Gamma\!_{;j}\, \bar\xi^j\!\!,_{i}.
\ee   
Using this to replace $\Gamma\!,_i$ in the mspWI, say \eqref{mspwi_2}, gives
\be \label{mspwi_cov1}
\tilde\Gamma\!,_{i} +\Gamma\!_{;j}(\langle\xi^j\!\!,_{i} \rangle - \bar\xi^j\!\!,_{i}) -{\textstyle{\frac{1}{2}}} G^{mn}(R_{nm}),_{i} -G^{np} R_{pm}\langle \xi^m\!\!\!,_i \rangle_{;n} =0.
\ee
The first term is now a covariant vector as already mentioned. In fact, the second term is also covariant.  This can be made manifest by replacing the ordinary background derivatives by covariant background derivatives 
\be 
\Gamma\!_{;j}(\langle\xi^j\!\!,_{i} \rangle - \bar\xi^j\!\!,_{i}) = \Gamma\!_{;j}(\langle \nabla_{\!i}\xi^j \rangle - \nabla_{\!i}\bar\xi^j).
\ee
This works, of course, for any covariant derivative. The two extra terms proportional to the Christoffel symbols cancel out in this expression. A similar cancellation occurs when replacing the ordinary derivatives by covariant derivatives in the last two terms proportional to the cutoff
\be 
{\textstyle{\frac{1}{2}}} G^{mn}(R_{nm}),_{i} + G^{np} R_{pm}\langle \xi^m\!\!\!,_i \rangle_{;n} = {\textstyle{\frac{1}{2}}} G^{mn}\,\nabla_{\!i} R_{nm} +G^{np} R_{pm}\langle \nabla_{\!i} \xi^m \rangle_{;n}.
\ee
The manifestly covariant mspWI then takes the following form
\be \label{mspwi_cov}
\boxed{\mathcal{N}_i \equiv \tilde\Gamma\!,_{i} +\Gamma\!_{;j}(\langle \nabla_{\!i}\xi^j \rangle - \nabla_{\!i}\bar\xi^j) - {\textstyle{\frac{1}{2}}} G^{mn}\,\nabla_{\!i} R_{nm} -G^{np} R_{pm}\langle \nabla_{\!i} \xi^m \rangle_{;n} =0}
\ee
In subsection \ref{consistency} we found the flow equation \eqref{N} for the quantity $\mathcal{N}_i$ above. We had further shown that the two pieces $\mathcal{N}_{1i}$ and $\mathcal{N}_{2i}$ follow the same flow equation, and this was derived without specifying the field space parametrization. Consequently, one expects that the same flow equation \eqref{N} holds for the quantities $\mathcal{N}_{1i}$ and $\mathcal{N}_{2i}$ after making them covariant by replacing $\partial\rightarrow\nabla$. This is indeed true as one can easily check. In fact the replacement $\partial\rightarrow\nabla$ in $\mathcal{N}_{1i}$ introduces some extra terms  which (cancel those of $\mathcal{N}_{2i}$ and) satisfy the flow \eqref{N}. In summary, the covariant versions of $\mathcal{N}_{1i}$ and $\mathcal{N}_{2i}$, denoted by a bar, are defined as
\be \label{N_12_cov}
\bar{\mathcal{N}}_{1i} \equiv  \tilde\Gamma\!,_{i} -\Gamma\!_{;j}\nabla_{\!i}\bar\xi^j - {\textstyle{\frac{1}{2}}} G^{mn}\,\nabla_{\!i} R_{nm}, \hspace{1cm}
\bar{\mathcal{N}}_{2i} \equiv  \Gamma\!_{;j}\langle \nabla_{\!i}\xi^j \rangle -G^{np} R_{pm}\langle \nabla_{\!i} \xi^m \rangle_{;n},
\ee 
and follow the usual equation
\be \label{N12_cov}
\partial_t \bar{\mathcal{N}}_{1i} =  - {\textstyle{\frac{1}{2}}}\, (G\dot R G)^{qp}(\bar{\mathcal{N}}_{1i})_{;pq}, \hspace{5mm} \partial_t \bar{\mathcal{N}}_{2i} =  - {\textstyle{\frac{1}{2}}}\, (G\dot R G)^{qp}(\bar{\mathcal{N}}_{2i})_{;pq}.
\ee 
As commented in the last paragraph of subsection \ref{short}, extra terms may arise in \eqref{mspwi_cov} when using measures other than the one with only total-field dependence. These terms are also expected to be covariant if the measure is so, as exemplified by the last term in \eqref{mspwi_5}. Clearly, such terms also satisfy the flow \eqref{N}, by the general equation \eqref{Odot}.

\section{Gauge theories} \label{gauge}
 
The formalism we have presented so far, applies to non-gauge theories and gauge theories before the gauge fixing procedure. Of course, after gauge fixing, we will end up with a non-gauge theory and expect our arguments to go through, but this is not a priori clear. In particular, as a necessary step in the process of gauge fixing, we need to show that the effective action is gauge invariant at all scales. We will adopt the covariant approach of the previous section which turns out to be a requirement for achieving a gauge invariant effective action. We review here briefly the geometric approach to gauge theories and refer the reader to the literature for more details \cite{dewitt_gaqft,kunstatter_1992,parker_toms}.  

\subsection{Geometry}

Let us assign the same symbols used for non-gauge theories, to the coordinates of the gauge theory field space $\phi^i$, as well as their decomposition into background $\varphi^i$ and dynamical fields $\xi^i$. We also take the vector fields $K_\alpha$ as a basis for the generators of the gauge group, which form a closed algebra, and denote their components, at the point $\phi$, by $K^i_\alpha[\phi] \equiv K_\alpha \phi^i$. As is true for Yang-Mills theory and gravity, we assume the existence of a metric $g_{ij}$ on the field space, which enables us to define the effective action in a covariant way. For this purpose, the dynamical fields $\xi^i$ are chosen to be vectors satisfying a geodesic equation, as detailed in the previous section. However, the connection $\nabla^V$ used here to define the geodesic equations is not chosen to match exactly the one compatible with the field space metric, but rather it is defined by the condition $\nabla^V_{\!k} g^{\perp}_{ij}=0$, where $g^{\perp}_{ij}= P^m_i P^n_j g_{mn}$ is the metric projected onto the space orthogonal to the orbits, by the projection operator $P^i_j \equiv \delta^i_j - K^i_\alpha \gamma^{\alpha\beta}K^k_\beta g_{kj}$, where $\gamma^{\alpha\beta}$ is the inverse of the metric $\gamma_{\alpha\beta} = g_{ij}K^i_\alpha K^j_\beta$, defined on the orbits. This is known as the Vilkovisky connection\footnote{This definition of the Vilkovisky connection is due to S.Carlip (see e.g. footnotes of \cite{kunstatter_1992,parker_toms}).}. The condition $\nabla^V_{\!k} g^{\perp}_{ij}=0$ does not fix the connection completely but only up to terms which are irrelevant for the construction of the effective \nolinebreak action.

The process of gauge fixing consists of choosing a surface $\mathscr{S}$ in field space which intersects the gauge orbits once and only once. One can then choose a set of coordinates which is adapted to this choice. This consists of parametrizing the orbits with a set of fields, which take the same value on $\mathscr{S}$, and assigning a set of coordinates to the surface $\mathscr{S}$. To avoid complicating the notation, the coordinates are chosen to be denoted by the same symbol used for a general coordinate system, but with the super index $i$ running over small Greek indices for the orbit parameters, $\phi^\alpha$, and taking capital Latin indices for the coordinates on $\mathscr{S}$, $\phi^I$, which label different orbits. We will therefore explicitly specify in the following whether we are using adapted or general coordinates. The adapted coordinates are of course not uniquely defined. The field redefinitions $\phi^I\rightarrow \phi'^I(\phi^I)$, and $\phi^\alpha\rightarrow \phi'^\alpha(\phi^\alpha)$ correspond to the same choice of gauge fixing condition, but provide a different parametrization for the adapted coordinate system, while the more general field redefinitions $\phi^I\rightarrow \phi'^I(\phi^I)$, and $\phi^\alpha\rightarrow \phi'^\alpha(\phi^I,\phi^\alpha)$ lead to some adapted coordinates with a different choice of gauge.

The definition of the Vilkovisky connection described above, is equivalent, in the adapted coordinates, to the following statement, in terms of the corresponding Christoffel symbols
\be \label{vc} 
(\Gamma_V)^K_{IJ} = {\textstyle{\frac{1}{2}}} h^{KL}(\partial_I h_{LJ}+\partial_J h_{LI}-\partial_L h_{IJ}), \hspace{1cm} (\Gamma_V)^K_{\alpha j} =0, \hspace{1cm} \partial_\alpha h_{IJ} =0,
\ee
where $h_{IJ}$ is the metric $g^{\perp}_{ij}$ induced on the gauge slice. An important consequence of this, which in fact motivates its definition, is that the component of the dynamical vector field along the orbit space, $\xi^I$, is independent of the orbit parameters $\varphi^\alpha$ and $\phi^\alpha$, and one can therefore write its functional dependence as $\xi^I(\varphi^I,\phi^I)$. This will be used repeatedly in the argument for gauge invariance. Let us also point out that in the adapted coordinates, $K^I_\alpha =0$ by construction, and the matrix $K^\beta_\alpha$ is assumed to be invertible. 

For completeness let us also sketch briefly how the divergence in the path integration over the redundant field space is removed, and refer the reader to \cite{parker_toms} for further details. The natural volume element which leads to a covariant effective action is 
\be 
\prod_i d\phi^i \sqrt{\mathrm{det}\, g_{ij}(\phi)}.
\ee
Using the decomposition of the line element $g_{ij}\,d\phi^i d\phi^j = g^\perp_{ij}\, d_\perp\!\phi^i d_\perp\!\phi^j +  \gamma_{\alpha\beta}\, d\epsilon^\alpha\! d\epsilon^\beta$, where $d_\perp\!\phi^i \equiv P^i_j d\phi^j$ and $d\epsilon^\alpha \equiv \gamma^{\alpha\beta} K^i_\beta g_{ij}d\phi^j$, the volume element is decomposed as
\be \label{vege} 
\prod_\alpha d\epsilon^\alpha \prod_i d_\perp \phi^i \sqrt{\mathrm{det}_\perp g^\perp_{ij}(\phi)} \sqrt{\mathrm{det}\,\gamma_{\alpha\beta}(\phi)},
\ee
where, by $\mathrm{det}_\perp$ the determinant in the space orthogonal to the orbits is meant. Written in the adapted coordinates, this takes a more transparent form
\be \label{vead} 
\prod_\alpha d\epsilon^\alpha \prod_I d \phi^I \sqrt{\mathrm{det}\, h_{IJ}(\phi^K)} \sqrt{\mathrm{det}\,\gamma_{\alpha\beta}(\phi^I)},
\ee
where, with abuse of notation, the same symbol $\gamma_{\alpha\beta}$ used in a general coordinate system in \eqref{vege} is used also here in the adapted coordinates. Apart from the leftmost factor, this expression depends only on the orbit space parameters $\phi^I$.

A gauge invariant integrand depends solely on $\phi^I$, and consequently the divergent integral over $\prod_\alpha d\epsilon^\alpha$ will drop out in expectation values of gauge invariant quantities. Therefore we are finally left with an integral over the orbit space. Notice that no ghost fields appear in this approach as the path integral is taken only over the equivalence classes of fields, or orbit space fields $\phi^I$, and not the whole redundant field space. The connection with the standard Faddeev-Popov method is made by introducing in the path integral over the orbit space, the measure $\prod_\alpha\! d\epsilon^\alpha\, \delta[\epsilon^\alpha]$ including a Dirac delta functional, whose integral equals unity, and changing variables back to the general coordinate system. This requires a Jacobian which gives rise to the Faddeev-Popov determinant.

\subsection{Gauge invariance} 

After this brief description of the geometry of gauge theories, we will now move on to discuss gauge invariance of the effective action. Although the discussion can well be presented in a general coordinate system, the steps of the argument will be more clearly conveyed when presented in the adapted coordinates. The covariant approach we have taken guarantees that there will be no loss of generality in doing so. The results will be finally restated in a coordinate-independent manner. From now on, we therefore take, with abuse of notation, the symbols $\phi^i$, used for a general coordinate system, to coincide with the adapted coordinates.  

We will put a tilde on the effective action when expressed as a function of the background and the total field $\bar\phi^i$, as in the previous section, and drop the bar on $\bar\phi^i$ and $\bar\xi^i$ from now on: $\tilde \Gamma[\varphi,\phi(\varphi,\xi)] = \Gamma[\varphi,\xi(\varphi,\phi)]$. With this notation, for a general functional $F$, invariance under gauge transformations of the total(background) field is equivalent to independence of $\phi^\alpha$($\varphi^\alpha$):
\be \label{gia} 
\frac{\delta \tilde F[\varphi,\phi]}{\delta \phi^\alpha} =0, \hspace{1cm} 
\frac{\delta \tilde F[\varphi,\phi]}{\delta \varphi^\alpha} =0.
\ee

The effective action is given as in \eqref{eaa_def} except that the generator of connected diagrams is defined by taking the path integral measure to be the determinant of the field space metric, evaluated at $\phi^i$, as dictated by the covariant formulation
\be  \label{wgt} 
\exp\left(-W_k[\varphi,J]\right) = \int \!\! D\phi \;\sqrt{\mathrm{det}\,g_{ij}(\phi)} \,\; \mathrm{exp}\left(-\tilde S[\phi^I]-S_k[\varphi,\xi]- J\!\cdot\!\xi\right), \hspace{5mm} 
\tilde S[\phi^I(\varphi,\xi)] = S[\varphi,\xi].
\ee
The gauge invariance of the ultraviolet action is emphasized by making the orbit index $I$ explicit in its argument. The ultraviolet action, therefore has the following properties
\be \label{gibts1} 
\tilde S,_{\alpha} =0, \hspace{1cm} \tilde S_{;\alpha} =0,
\ee
where, as noted after eq.\eqref{der}, here the ``$,$'' and ``$;$'' notations refer to derivatives with respect to the first and second arguments respectively.
Since $\phi^I$ is only a function of $\varphi^I$ and $\xi^I$, one can infer, from the right equation in \eqref{gibts1}, that the ultraviolet action is also independent of the components of the dynamical field along the gauge orbits, $\xi^\alpha$: $S_{;\alpha} =0$. In fact, given the $\varphi^\alpha$, $\phi^\alpha$ independence of $\xi^I$, the conditions \eqref{gibts1} are equivalent to their tilda-less versions
\be \label{gibts2} 
S,_{\alpha} =0, \hspace{1cm} S_{;\alpha} =0.
\ee
Now, let us consider the $k\rightarrow\infty$ limit of the effective action defined in this way. This can be shown to be
\be \label{kinfty}
\Gamma \rightarrow \tilde S[\phi^I] + {\textstyle{\frac{1}{2}}} \mathrm{Tr}\log R_{ij} - {\textstyle{\frac{d}{2}}}\log(2\pi) + {\textstyle{\frac{1}{2}}} \mathrm{Tr}\log g_{ij} + \mathrm{Tr}\log\, (\delta\xi^i/\delta\phi^j),
\ee
where $d$ is the spacetime dimension. The last two terms vanish in a regularization scheme, such as dimensional regularization, where $\delta(0)=0$. So in such a regularization scheme it is clear 
that in the large $k$ limit  
\be \label{uvc1} 
\Gamma\!,_{\alpha} - {\textstyle{\frac{1}{2}}} G R,_{\alpha} \rightarrow (S + {\textstyle{\frac{1}{2}}} \mathrm{Tr}\log R_{ij}),_{\alpha} - {\textstyle{\frac{1}{2}}} R^{-1} R,_{\alpha} = S,_\alpha = 0,
\ee
\be \label{uvc2} 
\Gamma_{;\alpha} \rightarrow S_{;\alpha} =0.
\ee
In fact, the second term in \eqref{kinfty} also vanishes with our choice of regularization, but it cancels anyway with the term $\frac{1}{2} G R,_{\alpha}$ in the expression \eqref{uvc1}. Then, from the left equation in \eqref{N12} it follows that the identity \eqref{uvc1} is valid at all scales. On the other hand, \eqref{uvc2} together with \eqref{gamma_dot;i}, implies that $\Gamma_{;\beta}=0$ at all scales. In summary, at all scales we have
\be \label{gigamma}
\Gamma_{\!,\alpha} - {\textstyle{\frac{1}{2}}} G R_{,\alpha}  =0 \hspace{1cm}
\Gamma_{;\alpha}=0.
\ee
This implies that the effective action depends on $\varphi^I$, $\varphi^\alpha$ and also $\xi^I$, which is itself a function of $\varphi^I,\phi^I$: $\Gamma[\varphi^I,\varphi^\alpha,\xi^I(\varphi^I,\phi^I)]$. This means that the effective action, being independent of $\phi^\alpha$, is invariant under gauge transformations of the total field. The extra $\varphi^\alpha$ dependence goes away if there is no such dependence in the cutoff, $R_{,\alpha} =0$, implying $\Gamma,_{\alpha}=0$, $\Gamma_{;\alpha}=0$, which is, as already mentioned, equivalent to $\tilde\Gamma,_{\alpha}=0$, $\tilde\Gamma_{;\alpha}=0$. The effective action will therefore be invariant under gauge transformations of both the background and the total field.

For a gauge invariant ultraviolet action, the $\alpha$ component of the mspWI is automatically satisfied, while the $I$ component gives a nontrivial constraint. If we further assume that they are only the components of the dynamical field along the surface $\mathscr{S}$, $\xi^I$, whose momentum modes are cut off, or in other words, if the only nonzero components of the cutoff kernel are $R_{IJ}$, then the $I$ component of the mspWI is cast to exactly the same form as the original identity \eqref{mspwi_2}, but on the surface $\mathscr{S}$:
\be \label{mspwi_gauge}
\Gamma\!,_{I} -{\textstyle{\frac{1}{2}}} G^{MN}(R_{NM}),_{I} +\Gamma\!_{;J}\langle\xi^J\!\!,_{I} \rangle -G^{NP} R_{PM}\langle \xi^M\!\!,_I \rangle_{;N} =0.
\ee
In this case, also the inverse propagator has nonzero components only along the surface $\mathscr{S}$, $G_{\alpha i} =0$. This equation depends on $\varphi^\alpha$ unless the additional condition $R,_\alpha =0$ is assumed.

One may now ask how the conditions \eqref{gigamma} will look like in a general coordinate system. Using the fact that in the adapted coordinates $K^I_\alpha =0$ and that $K^\beta_\alpha$ is invertible, the equation on the right-hand side in \eqref{gigamma} can be written as 
\be 
K^i_\alpha[\varphi]\Gamma_{;i}=0,
\ee
which is covariant and will take the same form in all coordinates. However, the equation on the left-hand side doesn't seem to be covariant at first sight. Using \eqref{gammatilde} one can replace $\Gamma_{\!,\alpha}$ with 
$\tilde \Gamma\!,_\alpha$. This is because the term $\Gamma\!_{;j}\, \xi^j\!\!,_\alpha = \Gamma\!_{;\beta}\, \xi^\beta\!\!,_\alpha+\Gamma\!_{;J}\,\xi^J\!\!\!,_\alpha$ vanishes by the fact that $\Gamma\!_{;\beta}=0$ and $\xi^J\!\!\!,_\alpha=0$. One can also replace the ordinary derivative of the cutoff with a covariant derivative $\nabla^{V}$, computed with the Vilkovisky connection. The extra terms involving the Christoffel symbols can be shown to vanish by our regularization choice and the property $(\Gamma_V)^K_{\alpha j} =0$ of the Vilkovisky connection \eqref{vc}. The condition can then be rewritten as 
\be \label{bgi_cov}
K^i_\alpha[\varphi](\tilde \Gamma_{\!,i} 
- {\textstyle{\frac{1}{2}}} G\,\nabla^{V}_{\!i}\! R)  =0. 
\ee
This is now written in a covariant way and will take the same form in any coordinate system. From this equation it is seen that $K^k_\alpha[\varphi]\nabla^{V}_{\!k}\! R_{ij}=0$ implies background gauge invariance of the effective action, as expected also from the equivalent condition in the adapted coordinates. Also the covariant form of the condition $R_{\alpha i} =0$ is $K^i_\alpha[\varphi]R_{ij}=0$. It is worth mentioning that, given $K^i_\alpha[\varphi]R_{ij}=0$,  the condition for background gauge invariance $K^k_\alpha[\varphi]\nabla^{V}_{\!k}\! R_{ij}=0$ is equivalent to the vanishing of the cutoff under Lie derivatives with respect to the gauge group generators (evaluated at the background point) $\mathcal{L}_{\scriptsize{K_\alpha}}\! R_{ij}=0$. 

The quantity on the left-hand side of \eqref{bgi_cov}, although covariant, doesn't seem to completely match the expression $\bar{\mathcal{N}}_{1i}$ defined at the end of the previous section, because it lacks the second term in $\bar{\mathcal{N}}_{1i}$. But this term actually vanishes $K^i_\alpha[\varphi]\Gamma\!_{;j}\nabla^{V}_{\!i}\!\xi^j=0$ by gauge invariance and the properties of the Vilkovisky connection. As a result, this quantity follows the usual flow equation \eqref{N}, as expected.

In summary, in a general coordinate system, gauge invariance of the ultraviolet action implies that $K^i_\alpha[\varphi]\Gamma_{;i}=0$, or equivalently $K^i_\alpha[\phi]\tilde\Gamma_{;i}=0$, which is the covariant version of invariance under gauge transformations of the total field. This fact, together with the assumption $K^i_\alpha[\varphi]R_{ij}=0$ implies that, in the adapted coordinates, the effective action satisfies the usual mspWI on the surface $\mathscr{S}$. On the other hand, background gauge invariance $K^i_\alpha[\varphi]\tilde \Gamma_{\!,i}=0$, or equivalently $K^i_\alpha[\varphi]\Gamma_{\!,i}=0$, follows with the additional condition $\nabla^{V}_{\!\alpha}\! R_{ij} =0$. Along with total-field gauge invariance, this tells us that the effective action is only a function of the coordinates on $\mathscr{S}$, in which case (6.10) will also be covariant under $\phi^I \rightarrow \phi'^I(\phi^I)$.

Changing the gauge fixing condition is equivalent to applying a field redefinition (in the adapted coordinates) of the form $\phi^\alpha\rightarrow \phi'^\alpha(\phi^I,\phi^\alpha)$. This transformation of the fields does not affect the effective action, simply because it is, by construction, a scalar under general coordinate transformations on all the field space, and not only the surface $\mathscr{S}$, and because in the adapted coordinates, the effective action is independent of the fields $\varphi^\alpha$ and $\phi^\alpha$ by gauge invariance.

\section{Conclusions}

In a quantum field theory with an infrared regulator and within the background-field framework, we have introduced the notion of splitting symmetry in its most general sense, and provided a simple and general path integral derivation of its Ward identity, which we have referred to as the mspWI.  

We have shown that the quantity $\mathcal{N}_i$ whose vanishing gives the Ward identity, can be divided, as in \eqref{N_12}, into two pieces $\mathcal{N}_{1i}$, $\mathcal{N}_{2i}$, each of which follows a simple flow equation \eqref{N12}. This proves crucial in finding the condition for background gauge invariance, at an arbitrary energy scale, in a geometric approach to gauge theories. 

The mspWI for the effective action, encompasses the information from the single-field dependence of the ultraviolet action. In particular, in the infrared limit, this implies that the effective action is also a functional of a single field $\Phi$, defined implicitly in \eqref{bargamma}. A redefinition $\xi\rightarrow\Phi$ therefore absorbs the second term in \eqref{mspwi_2} and makes the terms responsible for background dependence manifest.    

For the special case of exponential splitting, which results in a covariant effective action, we have shown that the mspWI is also covariant, i.e. that the structure of the mspWI does not change under field redefinitions. Furthermore, we have discussed the covariance of the flow equations \eqref{N12}. As the derivation suggests, these flow equations, although not manifestly covariant, are valid in any coordinate system. In fact in the non-covariant quantity $\mathcal{N}_{1i}$, the extra terms arising as a result of a field redefinition satisfy a similar flow equation separately. This can be put in a different way: that the covariantized versions of $\mathcal{N}_{1i}$ and $\mathcal{N}_{2i}$, while summing up to $\mathcal{N}_i$, also satisfy the usual flow \eqref{N12}.

The effective action can be computed perturbatively and is expected to satisfy the mspWI order by order in a loop expansion. This is explicitly verified to be the case at the one-loop level irrespective of the scheme of regularization. In performing this check, we have emphasized the unavoidable role of the path integral measure chosen in \eqref{W}. 

It is argued that the mspWI is generically divergent. This prevents the use of this identity in practice, to constrain the effective action, except in special cases such as the linear split. To overcome this problem, one needs to deal with the renormalized mspWI. For this purpose, we have introduced the modified master equation for the splitting symmetry, and with its aid, discussed how for theories renormalizable in its modern sense, the mspWI can be renormalized in the presence of the regulator, and that the renormalized master equation has the same structure as its unrenormalized counterpart. 

The Vilkovisky-DeWitt construction for general gauge theories is presented in the renormalization group context of \cite{wetterich_rg,morris_rg}. It is shown that the effective action is invariant under gauge transformations of the total field, and using \eqref{N12}, the condition for background gauge invariance is found. In particular, this provides as a special case a nonperturbative proof of gauge invariance of the infrared effective action. This is seen by simply setting k=0 in \eqref{bgi_cov}. In this particular argument, even if one is not interested in the scale dependence of the effective action but only its infrared limit, the regulator can be regarded merely as a tool, introduced at an intermediate step of the proof, to connect the ultraviolet action and the infrared effective action, and using the simple and exact flows \eqref{N12}, to transfer the information of gauge invariance from the ultraviolet to the infrared. Finally, provided that the cutoff does not have any components along the gauge orbits, the mspWI holds in its original form \eqref{mspwi_2}, also on the gauge fixing surface. 

The formalism presented in this work is expected to have important consequences for functional renormalization group applications to quantum field theories with background fields. In particular, in the renormalization group approach of \cite{wetterich_rg,morris_rg}, the covariant formulation reviewed in section \ref{cef} has been previously applied to nonlinear sigma models \cite{percacci_0810,percacci_0910,percacci_1010,percacci_1102,percacci_1105,zanusso_1207,safari_1306,safari_1406} and has more recently received attention in applications to gravity \cite{eichhorn_1301,nink_1410,codello_1412,percacci_1501,falls_1503,percacci_1505,percacci_1506,percacci_1507,gies_1507} (see also \cite{nink_1506}). Just as in the case of linear split \cite{morris_1312,morris_1502}, the mspWI for exponential splitting is an essential ingredient for finding consistent truncations in such studies. 

\section*{Acknowledgments}

I would like to thank Roberto Percacci for reading and checking parts of an early draft. I am also grateful to Gaurav Narain, Roberto Percacci, and Omar Zanusso for useful discussions and comments. Special thanks to Omar Zanusso for encouragement to complete this note and make it public. I also thank Tim Morris and Jan Pawlowski for useful comments and feedback on the draft.

\appendix

\section{Feynman rules} \label{feyn}

In this appendix we introduce some Feynman rules which are used in subsection \ref{short} to write the mspWI in diagrammatic language. There are three types of vertices which appear in the mspWI. \nopagebreak[4] These are shown by the first three of the diagrams below  
{\setlength\arraycolsep{2pt}
\bea 
\raisebox{-1cm}{\includegraphics[trim=7cm 21cm 7cm 2cm, clip=true, width=0.14\textwidth]{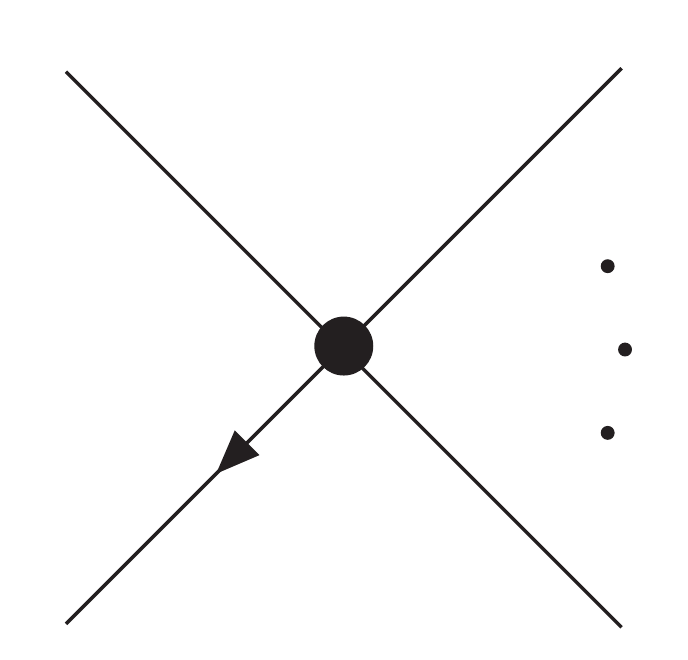}}
\hspace{-1mm}\raisebox{-10mm}{$i_n$}\hspace{1mm}
\hspace{-4mm}\raisebox{10mm}{$i_2$}\hspace{4mm}
\hspace{-30mm}\raisebox{-10mm}{$j$}\hspace{30mm}
\hspace{-32mm}\raisebox{10mm}{$i_1$} \hspace{2.71cm} &=& -\,R_{jk}\, C^k_{i,\,i_1i_2\cdots i_n}, 
\hspace{8mm} n\geq 1, \label{v1} \\[4mm]
\raisebox{-13mm}{\includegraphics[trim=7cm 20cm 7cm 2cm, clip=true, width=0.14\textwidth]{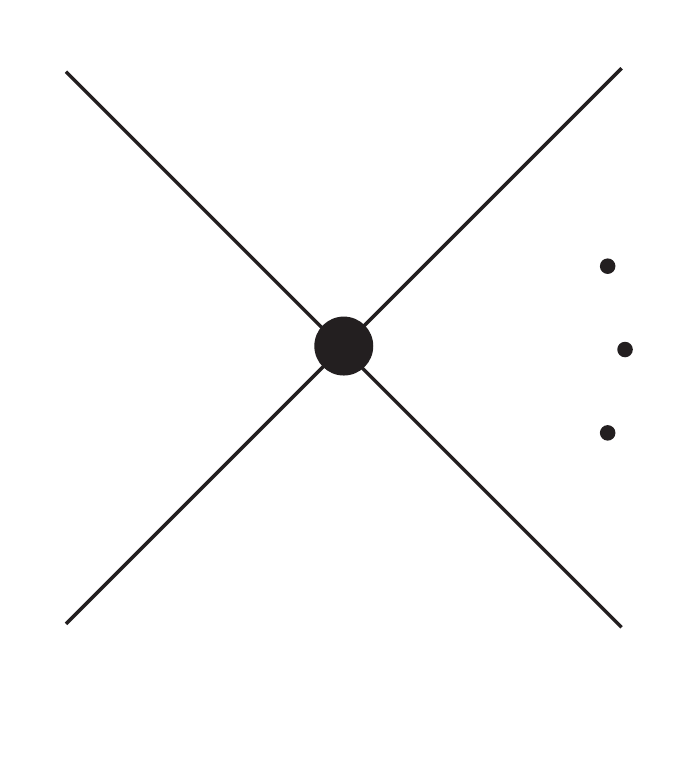}}
\hspace{-1mm}\raisebox{-10mm}{$i_n$}\hspace{1mm}
\hspace{-4mm}\raisebox{10mm}{$i_3$}\hspace{4mm}
\hspace{-30mm}\raisebox{-10mm}{$i_1$}\hspace{30mm}
\hspace{-33mm}\raisebox{10mm}{$i_2$} \hspace{2.72cm} &=& \hspace{2.5mm} \Gamma_{\!;j}\,C^j_{i,\,i_1i_2\cdots i_n}, \hspace{8mm} n\geq 0, \label{v2} \\[3mm]
\raisebox{-13mm}{\includegraphics[trim=7cm 20cm 7cm 2cm, clip=true, width=0.14\textwidth]{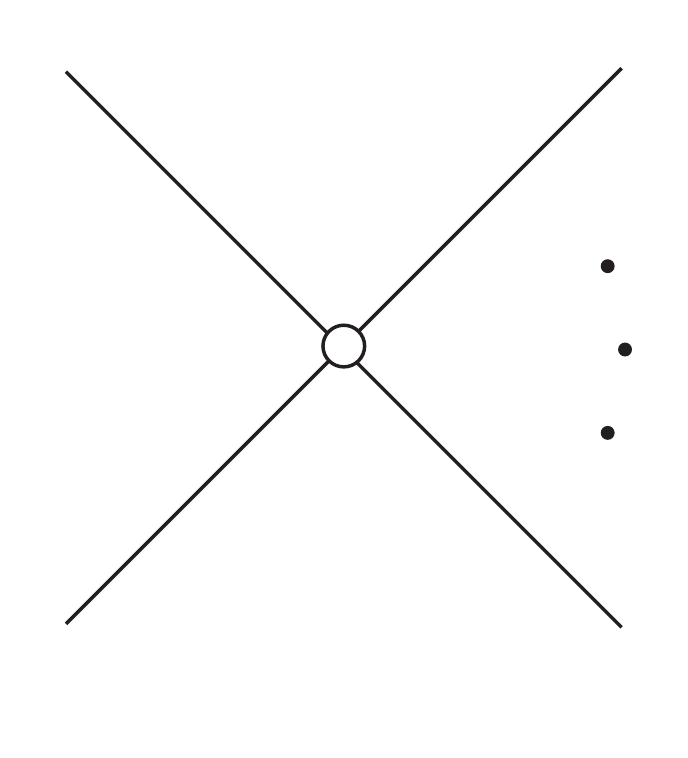}}
\hspace{-1mm}\raisebox{-10mm}{$i_n$}\hspace{1mm}
\hspace{-4mm}\raisebox{10mm}{$i_3$}\hspace{4mm}
\hspace{-30mm}\raisebox{-10mm}{$i_1$}\hspace{30mm}
\hspace{-33mm}\raisebox{10mm}{$i_2$} \hspace{2.72cm} &=& \hspace{3mm} \langle \xi^{i_1}\xi^{i_2}\cdots\xi^{i_n}\rangle_c, \hspace{1.2cm} n\geq 1, \label{v3} \\[1mm]
\raisebox{-13mm}{\includegraphics[trim=7cm 20cm 7cm 2cm, clip=true, width=0.14\textwidth]{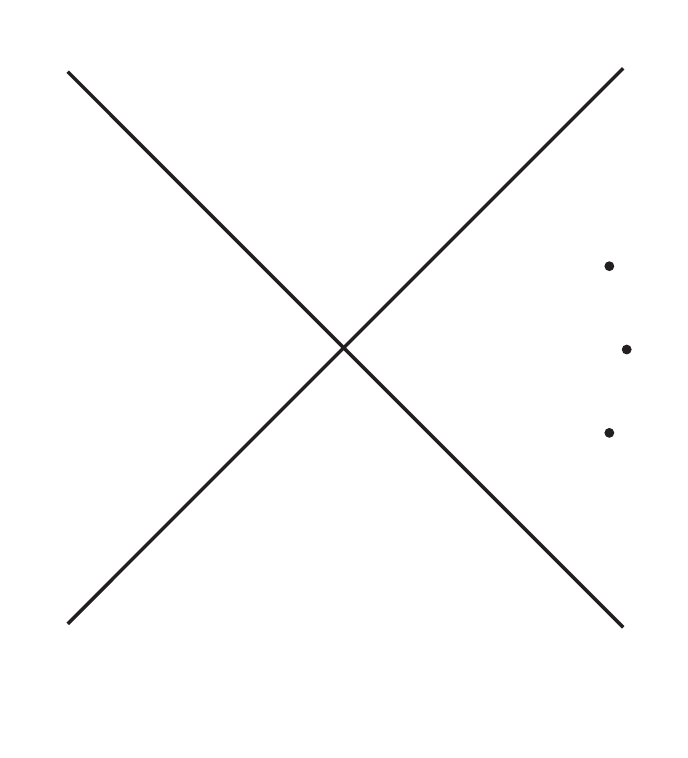}}
\hspace{-1mm}\raisebox{-10mm}{$i_n$}\hspace{1mm}
\hspace{-4mm}\raisebox{10mm}{$i_3$}\hspace{4mm}
\hspace{-30mm}\raisebox{-10mm}{$i_1$}\hspace{30mm}
\hspace{-33mm}\raisebox{10mm}{$i_2$} \hspace{2.72cm} &=& \; -\,\Gamma_{\! ; i_1 i_2\cdots i_n}, \hspace{1.2cm} n\geq 3. \label{v4} 
\eea}%
There is an $i$ index implicit in the first two vertices \eqref{v1} and \eqref{v2} with a black circle. The arrow in \eqref{v1} represents the free index on the cutoff and shouldn't be confused with momentum flow. The vertices with a white circle \eqref{v3} denote connected $n$-point functions, whose dependence on the background $\varphi^i$ and fluctuation field $\xi^i$ is expressed in a more explicit way when written in terms of one-particle irreducible vertices \eqref{v4} and the propagators (\eqref{v3} with $n=2$). For instance, the connected three and four-point functions are expressed in terms of the one-particle irreducible vertices as 
\be 
\raisebox{-7mm}{\includegraphics[trim=4cm 22cm 10cm 1cm, clip=true, width=0.14\textwidth]{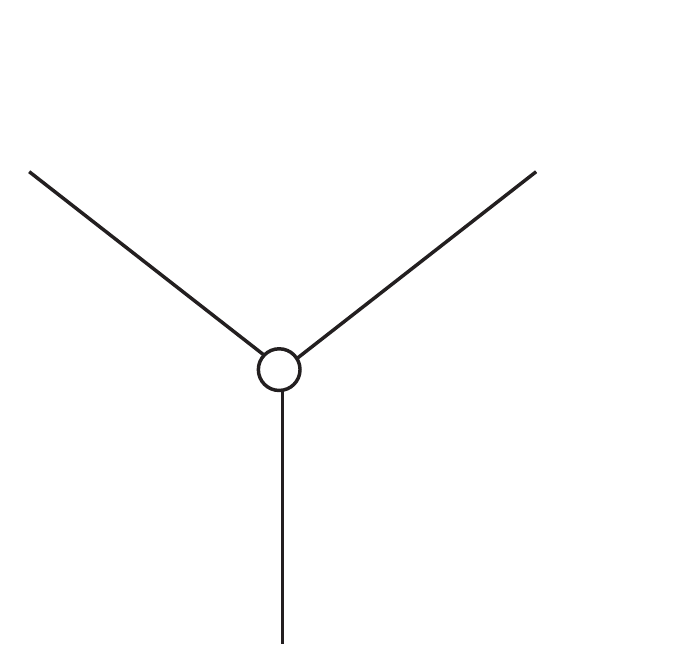}} =
\raisebox{-7mm}{\includegraphics[trim=10cm 22cm 4cm 1cm, clip=true, width=0.14\textwidth]{exp2}}
\ee 
\be 
\raisebox{-11mm}{\includegraphics[trim=3cm 20cm 10cm 1cm, clip=true, width=0.14\textwidth]{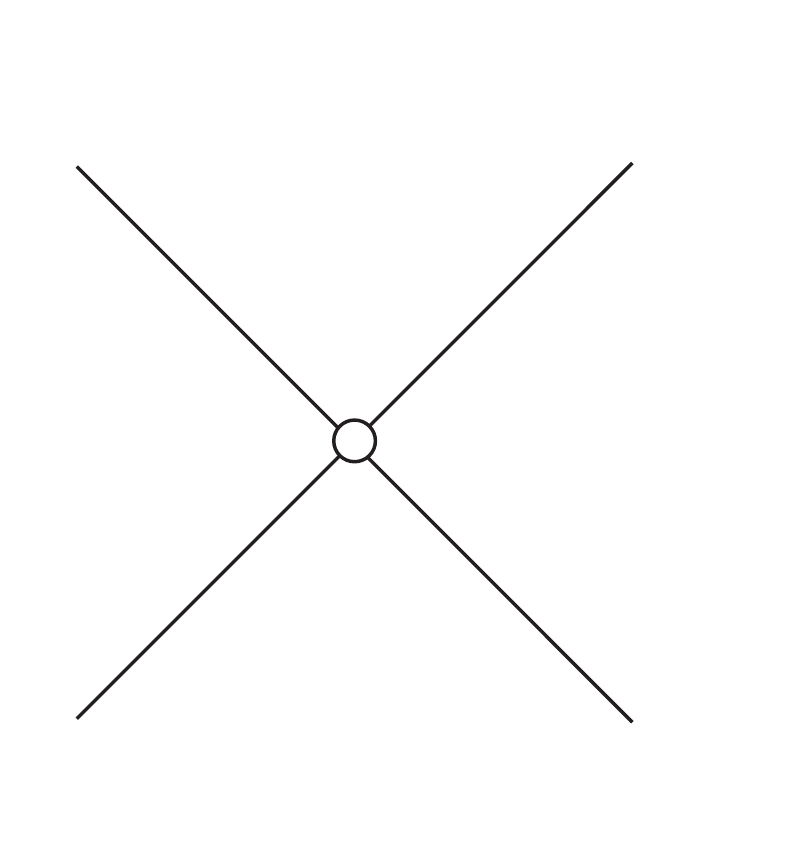}} =
\raisebox{-11mm}{\includegraphics[trim=10cm 20cm 3cm 1cm, clip=true, width=0.14\textwidth]{exp1}} + \left(\raisebox{-7mm}{\includegraphics[trim=3cm 14cm 6cm 10cm, clip=true, width=0.21\textwidth]{exp1}} + \;\;\mathrm{two\;\,other\;\,permutations}\right)
\ee 
The convention of putting a minus sign in \eqref{v4} makes all the terms on the right-hand side of the above and higher order equations to appear with a plus sign. 

\section{Functional flows} \label{flows}

We provide here some important identities and flow equations used throughout the text. We start with the flow equation for the effective action \eqref{frge} in the compact notation
\be \label{gamma_dot}
\partial_t \Gamma = {\textstyle{\frac{1}{2}}} G^{mn}\dot R_{nm}.
\ee
The flow of the one-particle irreducible one-point function follows trivially by differentiating this equation with respect to the classical fluctuating field $\bar\xi^i$ 
\be \label{gamma_dot;i}
\partial_t \Gamma_{;i} = - {\textstyle{\frac{1}{2}}} G^{mp}\,\Gamma_{;ipq}\,G^{qn}\dot R_{nm} = - {\textstyle{\frac{1}{2}}} G^{qn}\dot R_{nm}G^{mp} \Gamma_{;ipq} =  - {\textstyle{\frac{1}{2}}} (G\dot R G)^{qp} \Gamma_{;ipq}.
\ee
Taking the second derivative we arrive at
{\setlength\arraycolsep{2pt}
\bea  
\partial_t \Gamma_{;ij} 
&=& - {\textstyle{\frac{1}{2}}} (G\dot R G)^{qp} \Gamma_{;ijpq} +  {\textstyle{\frac{1}{2}}} G^{qm} \Gamma_{;jmn} G^{nr} \dot R_{rs} G^{sp} \Gamma_{;ipq} + {\textstyle{\frac{1}{2}}} G^{qm}\dot R_{mn} G^{nr} \Gamma_{;jrs}   G^{sp} \Gamma_{;ipq} \nn\\
&=& - {\textstyle{\frac{1}{2}}} (G\dot R G)^{qp} \Gamma_{;ijpq} + (G\dot R G)^{pq} G^{mn} \Gamma_{;ipm} \Gamma_{;jqn} =  - {\textstyle{\frac{1}{2}}} (G\dot R G)^{qp} \left[\Gamma_{;ijpq} + 2 G^{mn} \Gamma_{;ipm} \Gamma_{;jqn}\right]. \label{gamma_dot;ij}
\eea}%
It is easy to generalize this to one-particle irreducible n-point functions by successive differentiation with respect to $\bar\xi^i$. Another useful ingredient is the flow of the expectation value of an arbitrary operator $O$. This can be computed as follows by simply referring to the definition of expectation values, based on the weight and integral measure, in \eqref{W}
{\setlength\arraycolsep{2pt}
\bea  
\partial_t\langle O \rangle 
&=& -{\textstyle{\frac{1}{2}}}\langle \xi\!\cdot\!\dot R\!\cdot\!\xi\; O \rangle +{\textstyle{\frac{1}{2}}}\langle \xi\!\cdot\!\dot R\!\cdot\!\xi\rangle\langle O\rangle -\dot J\!\cdot\!(\langle\xi\;O\rangle - \bar\xi\langle O\rangle) + \langle \dot O \rangle \nn\\
&=& -{\textstyle{\frac{1}{2}}}\langle \xi\!\cdot\!\dot R\!\cdot\!\xi\; O \rangle +{\textstyle{\frac{1}{2}}}\langle \xi\!\cdot\!\dot R\!\cdot\!\xi\rangle\langle O\rangle +\delta\dot\Gamma\!/\delta\xi\!\cdot\!(\langle\xi\;O\rangle - \bar\xi\langle O\rangle)+ \bar\xi\!\cdot\!\dot R\!\cdot\!\langle\xi\;O\rangle - \bar\xi\!\cdot\!\dot R\!\cdot\!\bar\xi\,\langle O\rangle + \langle \dot O \rangle \nn\\
&=& -{\textstyle{\frac{1}{2}}}\langle \xi\!\cdot\!\dot R\!\cdot\!\xi\; O \rangle +{\textstyle{\frac{1}{2}}}(\langle \xi\!\cdot\!\dot R\!\cdot\!\xi\rangle -\bar\xi\!\cdot\!\dot R\!\cdot\!\bar\xi)\,\langle O\rangle +\delta\dot\Gamma\!/\delta\xi\!\cdot\!(\langle\xi\;O\rangle - \bar\xi\langle O\rangle)+ \xi\!\cdot\!\dot R\!\cdot\!\langle\xi\;O\rangle - {\textstyle{\frac{1}{2}}}\bar\xi\!\cdot\!\dot R\!\cdot\!\bar\xi\,\langle O\rangle + \langle \dot O \rangle \nn\\
&=& -{\textstyle{\frac{1}{2}}}\langle \xi\!\cdot\!\dot R\!\cdot\!\xi\; O \rangle +{\textstyle{\frac{1}{2}}}G^{ij}\dot R_{ji}\,\langle O\rangle +\delta\dot\Gamma\!/\delta\xi\!\cdot\!(\langle\xi\;O\rangle - \bar\xi\langle O\rangle)+ \xi\!\cdot\!\dot R\!\cdot\!\langle\xi\;O\rangle - {\textstyle{\frac{1}{2}}}\bar\xi\!\cdot\!\dot R\!\cdot\!\bar\xi\,\langle O\rangle + \langle \dot O \rangle,
\eea}%
where in the second line we have used the middle equation in \eqref{step2}. The first and second $\xi$-derivatives of $\langle O\rangle$ are also easily computed
\be  \label{O;i}
\langle O \rangle_{;i} 
= -(\delta J_{\!j}/\delta\xi^i)\,\langle \xi^j O \rangle + (\delta J_{\!j}/\delta\xi^i)\,\bar\xi^j\,\langle O \rangle  = G_{ij}(\langle \xi^j O \rangle - \bar \xi^j\langle O \rangle)
\ee
\be \label{O;ij}
\langle O \rangle_{;ij} 
= \Gamma_{;ijk}(\langle \xi^k O \rangle - \bar\xi^k\langle O \rangle) - G_{ij}\langle O \rangle + G_{im}G_{jn}(\langle \xi^m\xi^n O \rangle -\bar\xi^m\langle \xi^n O \rangle -\bar\xi^n\langle \xi^m O \rangle +\bar\xi^m\bar\xi^n\langle O \rangle).
\ee
Comparing \eqref{Odot} and \eqref{O;ij} makes it tempting to compute the contraction
{\setlength\arraycolsep{2pt}
\bea  
\hspace{-5mm}-{\textstyle{\frac{1}{2}}} (G\dot R G)^{ij} \langle O \rangle_{;ij} 
&=& -{\textstyle{\frac{1}{2}}} (G\dot R G)^{ij}\Gamma_{;ijk}(\langle \xi^k O \rangle - \bar\xi^k\langle O \rangle)+{\textstyle{\frac{1}{2}}} G^{ij}\dot R_{ji}\langle O \rangle \nn\\
&& -{\textstyle{\frac{1}{2}}} \dot R_{mn}(\langle \xi^m\xi^n O \rangle -\bar\xi^m\langle \xi^n O \rangle -\bar\xi^n\langle \xi^m O \rangle + \bar\xi^m\bar\xi^n\langle O \rangle) \nn\\[1mm]
&=&  \dot\Gamma_{;k}(\langle \xi^k O \rangle \!- \bar\xi^k\langle O \rangle)+{\textstyle{\frac{1}{2}}} G^{ij}\dot R_{ji}\langle O \rangle -{\textstyle{\frac{1}{2}}} (\langle \xi\!\cdot\!R\!\cdot\!\xi O \rangle \!-2\bar\xi\!\cdot\!R\!\cdot\!\langle \xi O \rangle + \bar\xi\!\cdot\!R\!\cdot\!\bar\xi\langle O \rangle),
\eea}%
with the aid of which the desired flow equation follows
\be \label{Odot}
\boxed{\partial_t\langle O \rangle  = \langle \dot O \rangle -{\textstyle{\frac{1}{2}}}(G\dot R G)^{ij} \langle O \rangle_{;ij}}
\ee
The second term on the right-hand side can therefore be interpreted as the commutator of $t$-differentiation and the averaging process, acting on $O$. 

\section{Explicit formulas for the exponential parametrization} \label{exp}

In section \ref{cef} we emphasized the importance of the exponential parametrization. Once the way the total field is split is specified, one can find an explicit formula for $\delta\xi^i$ or equivalently for $\xi^i\!\!,_j$ in \eqref{dxi}. For this purpose, we need to refer to the explicit expression for the total field in terms of the background and the dynamical field
\be \label{phi_(varphi_xi)}
\phi^i=\varphi^{i}+\xi^{i} -\sum_{n=2}^{\infty}\frac{1}{n!}\, \Gamma^{i}_{i_1 i_2 \ldots i_n}(\varphi)\,\xi^{i_1}\cdots\xi^{i_n}, \hspace{1cm}  \Gamma^{i}_{i_1 i_2 \ldots i_{n+1}} \equiv  \nabla_{i_1}\Gamma^{i}_{i_2 \ldots i_{n+1}}, \hspace{5mm} n\geq 2,
\ee
where the covariant derivative is defined with the connection $\Gamma^k_{ij}$ itself, and is taken with respect to lower indices only.

If we now make a variation $\delta\varphi^i$ in the background field, the variation $\delta\xi^i$ in the fluctuations must be made in such a way as to leave $\phi^i$ untouched. So taking the derivative of \eqref{phi_(varphi_xi)} with respect to $\varphi^i$, keeping $\phi^i$ fixed, we find
\be \label{dphi=0}
0=\delta^i_j+ \xi^i\!\!,_j -\sum_{n=2}^{\infty}\frac{1}{n!}\, \Gamma^{i}_{i_1 i_2 \ldots i_n}(\varphi)\,\partial_j(\xi^{i_1}\xi^{i_2}\cdots\xi^{i_n})-\sum_{n=2}^{\infty}\frac{1}{n!}\,\partial_j\Gamma^{i}_{i_1 i_2 \ldots i_n}(\varphi)\,\xi^{i_1}\cdots\xi^{i_n}.
\ee
This identity is valid in any coordinate system, in particular in normal coordinates the second term can be replaced with $\nabla_{\!j}\xi^i$ and the third term vanishes, so it simplifies to 
\be \label{dphi=0_normal}
0 \stackrel{*}{=} \delta^i_j+\nabla_{\!j}\xi^i -\sum_{n=2}^{\infty}\frac{1}{n!}\,\partial_j\Gamma^{i}_{i_1 i_2 \ldots i_n}(\varphi)\,\xi^{i_1}\cdots\xi^{i_n},
\ee
where
\be  
\partial_j\Gamma^{i}_{i_1 i_2}(\varphi) \stackrel{*}{=} \frac{2}{3}\Riem{j}{(i_1}{i}{i_2)}(\varphi) , \hspace{8mm} \partial_j\Gamma^{i}_{(i_1 i_2 i_3)}(\varphi) \stackrel{*}{=} \frac{1}{2}\nabla_{(i_1}\Riem{\vert j\vert}{i_2}{i}{i_3)}(\varphi) , \hspace{8mm} \cdots
\ee
and a star on the equation means that the identity is valid only in normal coordinates. Substituting these into \eqref{dphi=0_normal} we get a \textit{tensor} identity in normal coordinates, so it is valid in any coordinate system. We can therefore write
\be  \label{d_xi}
\xi^i\!\!,_j = -\delta^i_j - \Gamma^i_{jk}\,\xi^k + \frac{1}{3}\Riem{j}{k}{i}{l}\,\xi^k\xi^l + \frac{1}{12}\nabla_k\Riem{j}{l}{i}{n}\,\xi^k\xi^l\xi^n + \cdots
\ee
From this expression one can identify the first few coefficients in \eqref{dxi}: 
\be 
C^j_{i} = -\delta^j_i, \hspace{5mm} 
C^j_{i,m} =  - \Gamma^j_{im}, \hspace{5mm}
C^j_{i,mn} = \frac{1}{3}\Riem{i}{(m}{j}{n)}, \hspace{5mm}
C^j_{i,mnk} = \frac{1}{12}\nabla_{(m}\Riem{|i|}{n}{j}{k)}, \hspace{5mm} \cdots
\ee

\end{document}